\documentclass{emulateapj}
\usepackage{graphicx}
\usepackage{latexsym}
\usepackage{textcomp}
\usepackage{longtable}
\usepackage{multirow,booktabs}
\usepackage{amsfonts,amsmath,amssymb}
\usepackage{url}
\usepackage{hyperref}

\hypersetup{colorlinks=false,pdfborder={0 0 0}}

\usepackage{natbib}

\DeclareGraphicsExtensions{.pdf,.png}

\newcommand\fnurl[2]{%
\href{#2}{#1}\footnote{\url{#2}}%
}
\newcommand\A[0]{\mathbf{A}}
\newcommand\B[0]{\mathbf{B}}
\newcommand\C[0]{\mathbf{C}}
\newcommand\I[0]{\mathbf{I}}
\newcommand\F[0]{\mathbf{F}}
\newcommand\G[0]{\mathbf{G}}
\newcommand\GammaMat[0]{\boldsymbol{\Gamma}}
\newcommand\x[0]{\mathbf{x}}

\renewcommand\vec[1]{\mathbf{#1}}
\newcommand\al[0]{\boldsymbol{\alpha}}
\newcommand\asinh[0]{\operatorname{asinh}}

\newcommand\rhat[0]{\vec{\hat{r}}}
\newcommand\drho{\delta\rho}

\newcommand\sF{{\mathcal F}}
\newcommand\Sigcrit{\Sigma_{\rm{cr}}}

\newcommand\betahat{\hat{\beta}}
\newcommand\tauhat{\hat{\tau}}

\newcounter{Affiliation}
\newcommand{\aftext}[1]{\refstepcounter{Affiliation}\altaffilmark{\theAffiliation}#1}

\begin{document}

\title{Quantifying Environmental and Line-of-Sight Effects in Models of \\ Strong Gravitational Lens Systems}


\author{Curtis McCully\altaffilmark{\ref{af:LCOGT},\ref{af:UCSB},*} \and 
Charles R. Keeton\altaffilmark{\ref{af:Rutgers}} \and
Kenneth C. Wong\altaffilmark{\ref{af:ASIAA},\ref{af:NAOJ},$\dagger$} \and
Ann I. Zabludoff\altaffilmark{\ref{af:Arizona}}
}

\affil{
\aftext{Las Cumbres Observatory, 6740 Cortona Dr Suite 102, Goleta, CA 93117-5575, USA\label{af:LCOGT}}\\
\aftext{Department of Physics, University of California, Santa Barbara, CA 93106-9530, USA\label{af:UCSB}}\\
\aftext{Department of Physics and Astronomy, Rutgers, the State University of New Jersey, 136 Frelinghuysen Road, Piscataway, NJ 08854, USA.\label{af:Rutgers}}\\
\aftext{Institute of Astronomy and Astrophysics, Academia Sinica (ASIAA), P.O. Box 23-141, Taipei 10617, Taiwan
\label{af:ASIAA}}\\
\aftext{National Astronomical Observatory of Japan, 2-21-1 Osawa, Mitaka, Tokyo 181-8588, Japan\label{af:NAOJ}}\\
\aftext{Steward Observatory, University of Arizona, 933 North Cherry Avenue, Tucson, AZ 85721, USA\label{af:Arizona}}
}
\altaffiltext{*}{\email{}{cmccully@lco.global}}
\altaffiltext{$\dagger$}{EACOA Fellow}

\begin{abstract}
Matter near a gravitational lens galaxy or projected along the line of sight (LOS) can affect strong lensing observables by more than contemporary measurement errors. We simulate lens fields with realistic three-dimensional mass configurations (self-consistently including voids), and then fit mock lensing observables with increasingly complex lens models to quantify biases and uncertainties associated with different ways of treating the lens environment (ENV) and LOS. We identify the combination of mass, projected offset, and redshift that determines the importance of a perturbing galaxy for lensing. Foreground structures have a stronger effect on the lens potential than background structures, due to nonlinear effects in the foreground and downweighting in the background. There is dramatic variation in the net strength of ENV/LOS effects across different lens fields; modeling fields individually yields stronger priors for $H_0$ than ray tracing through $N$-body simulations. Models that ignore mass outside the lens yield poor fits and biased results. Adding external shear can account for tidal stretching from galaxies at redshifts $z \ge z_{\rm lens}$, but it requires corrections for external convergence and cannot reproduce nonlinear effects from foreground galaxies. Using the tidal approximation is reasonable for most perturbers as long as nonlinear redshift effects are included. Even then, the scatter in $H_0$ is limited by the lens profile degeneracy. Asymmetric image configurations produced by highly elliptical lens galaxies are less sensitive to the lens profile degeneracy, so they offer appealing targets for precision lensing analyses in future surveys like LSST and Euclid.
\end{abstract}

\maketitle
\section{Introduction}
\setcounter{footnote}{0}
Strong gravitational lensing is an important probe for many facets of cosmology. Analysis of strong lenses has led to constraints on the masses and properties of dark matter halos of galaxies \citep[e.g.,][]{Keeton98,Koopmans06,Barnabe09,Auger10,Treu06,Treu10_galaxies,Lagattuta12,Wong14}, substructure in galaxy halos \citep[e.g.,][]{Mao98,Metcalf01,Dalal02,Hezaveh16,Vegetti14}, and the Hubble constant, independent of the cosmic distance ladder \citep[e.g.,][]{Refsdal64,Keeton97_PG1115,Kochanek03,Saha06,Oguri07,Suyu10,Suyu13, Birrer16, Chen16, Bonvin16, Suyu16, Wong16}. Strong lensing may also be employed to constrain the properties of dark energy \citep[e.g.,][]{Turner90,Linder04,Linder11,Cao12,Treu13}.

In recent years, both the quantity and quality of strong lens data have improved. There are now roughly a hundred known lensed quasi-stellar objects (QSOs) (e.g.,\ \fnurl{CASTLeS}{http://www.cfa.harvard.edu/castles/}; \citealt{CLASS, SQLS}) and a comparable number of strongly lensed galaxies \citep[e.g.,][]{Bolton08,Cassowary}. These samples will increase dramatically in the near future with LSST and Euclid \citep[e.g.,][]{LSST, Coe09, Oguri10, Refregier10, Collett15}. Here, we focus on lensed QSOs because the compact source can vary rapidly enough to enable measurements of lens time delays (though much of this discussion also applies to strongly lensed supernovae; \citealt{Kelly15, Kelly16, Rodney16}). The relative positions and fluxes of lensed QSO images are routinely measured to high precision using the \textit{Hubble Space Telescope} \citep[HST; e.g,.][and references therein; CASTLeS Collaboration]{Lehar00,Sluse12}. Our understanding of gravitational lenses and the constraints they place on cosmology are no longer limited by observations, but by systematic uncertainties. 

One of the key issues is that lens galaxies are not isolated systems. As we enter the era of ``Precision Lensing'' (which we define by the goal of constraining the Hubble constant to $<\!\!1\%$), the perturbations in the lensing potential due to matter in the environment of the lens galaxy and projected along the line of sight (hereafter ENV/LOS) cannot be ignored \citep[e.g.,][]{Bar-Kana96,Momcheva06,Wong11, Collett16}.  \citet{Jaroszynski14} show that omitting ENV/LOS effects can often lead to unsuccessful fits, especially when time delay constraints are included. \citet{Collett16} argue that lensed QSOs are more likely to have an overdense LOS than the global population which can lead to a bias in the inferred Hubble constant from a sample of lensed QSOs.

To lowest order, using the ``tidal approximation,'' perturbing galaxies contribute ``external convergence,'' which produces extra (de-)magnification, and ``external shear,'' which stretches the images of the source, transforming circles to ellipses. External convergence due to the ENV/LOS is one of the dominant components of the uncertainty budget for current measurements of the Hubble constant \citep{Suyu12}.

Our goal for this work is to quantify the effects of mass outside the main lens galaxy on the measured cosmology,\footnote{Here we focus on the Hubble constant, but similar arguments apply to time-delay cosmography measurements that have been proposed to be used to measure dark energy \citep{Treu13}.} addressing the following questions:
which individual perturbing galaxies are the most important?
What is the range of ENV/LOS contributions?
What drives the bias and scatter in reproducing lensing observables from lens systems in realistic ENV/LOS mass distributions?
Which lens systems produce the strongest constraints on the Hubble constant?

To answer these questions, we build realistic matter distributions using extensive photometric and spectroscopic data for $\sim25$ strong lensing systems \citep{Momcheva06,Momcheva15,Williams06}. We embed a lens galaxy in these models and ray trace through the full three-dimensional matter distributions to generate mock lensing observables. We then fit these lensing observables with models that treat the ENV/LOS in increasingly accurate (but complex) ways,
to understand the precision and accuracy of the recovered (main) lens galaxy properties and $H_0$.

The simplest model (which we term ``Lens-Only'', Section \ref{sec:lensonly}) neglects mass outside the main lens altogether. This model serves as a control for our experiments, allowing us to quantify the overall importance of the ENV/LOS for constraining cosmology. This approach almost surely produces biased constraints on the Hubble constant, and so it is rarely used in studies that derive constraints on precision cosmology, but it still appears in some studies that focus on lens and source properties \citep[e.g.,][]{Calanog14, Hezaveh13}.

Next we add one new term to our model in order to account for tidal effects from matter in the lens plane. In this ``Lens+Shear'' model (Section \ref{sec:lens+shear}), the magnitude and direction of the external shear are treated as free parameters to be optimized for individual lens systems. Such a model has some freedom to adjust for the ENV/LOS, but it has some limitations as well. First, the Lens+Shear model neglects nonlinear effects that arise from having mass in multiple redshift planes along the LOS \citep[][]{McCully14,Jaroszynski12}. Also, this model treats the entire ENV/LOS in the tidal approximation: this neglects higher-order effects beyond shear, which may be significant for objects sufficiently close to the optical axis. Lens models also cannot directly constrain the external convergence because of the mass sheet degeneracy \citep{Falco85}.  To avoid biases in the derived cosmological parameters, corrections for external convergence must be applied using additional constraints such as weak lensing \citep{Nakajima09, Fadely10} or the number density of galaxies near the lens \citep{Suyu10, Suyu13, Collett13, Greene13, Rusu16, Wong16} calibrated by ray tracing through cosmological simulations \citep[e.g.,][]{Hilbert09}. However, these corrections are derived from statistical studies that may have limited applicability to individual lens environments. The shear derived from lens models also does not always match the shear calculated from modeling the ENV/LOS directly \citep{Wong11}.

The third type of model we consider explicitly accounts for mass in the environment and along the LOS including redshift effects, but only in the tidal approximation so we have called this the ``Lens+3D Tides'' model  (Section \ref{sec:3Dtides}). The difference between this model and the Lens+Shear model is that the shear is no longer included only in the main lens plane. Instead we use a set of tensors that account for multiple redshifts and propagate their nonlinear effects through the multi-plane lens equation (\citealt{McCully14}, hereafter M14). For this analysis, the 3D tidal tensors are fixed, but in principle they could also be allowed to vary in the fits. By comparing the recovered parameters in these Lens+3D Tides models to the Lens+Shear models, we can quantify the contribution of the nonlinear redshift effects to the cosmological error budget. We note that the convergence is explicitly included in our 3D tidal tensors, so adding an external convergence in post-processing is unnecessary.

Finally, we test full ``3D lens'' models that include some perturbing galaxies exactly, rather than in the tidal approximation (Section \ref{sec:3Dlens}). By comparing the recovered cosmological parameters to the models that only include the 3D tidal tensors, we can disentangle nonlinear redshift effects, and test the importance of higher-order terms for the most important perturbing galaxies (we will quantify perturbing galaxy importance in Section \ref{sec:Individual}). As we include more and more galaxies exactly, our 3D Lens models become indistinguishable from the input lens equation that was used to generate the lensing observables. By examining the recovered parameters from these models, we can quantify the fundamental limitations on measuring the Hubble constant using gravitational lens models.

 Section \ref{sec:methods} describes our methodology. We start by discussing how we build 3D mass distributions (Section \ref{sec:massmodels}), including a new treatment of voids. We then generate mock lensing observables (Section \ref{sec:observables}). Finally, we fit different types of lens models to those observables to test how well we can recover parameters relative to the input galaxy and cosmology (Section \ref{sec:lensmodels}). Section \ref{sec:results} then presents our results. In Section \ref{sec:Individual}, we run a set simulations following the full procedure in Section \ref{sec:methods}, but for a simplified mass distribution with a lens galaxy and a single peturbing galaxy. We use these results to gain intuition  about what makes a perturber important for lensing, and to derive a parameter to quantitatively rank order the contributions from individual perturbing galaxies. In Section \ref{sec:Environments}, we use that quantity to characterize the mass distributions of different lens fields, which are built from observational data following the methods in \ref{sec:massmodels}. In Section \ref{sec:fitting}, we then examine the parameters recovered from the different types of fitted models, comparing them to the input, ``true'' values. In four steps, we progressively add complexity to the models: (1) Lens-Only models that ignore the environment, (2) Lens+Shear models that attempt to fit the effective shear but include no redshift information, (3) Lens+3D Tides that include nonlinear redshift effects, and (4) 3D Lens models that include higher-order terms for the ENV/LOS so we can test for fundemental limitations in gravitational lens modeling. Finally, in Section \ref{sec:ImageConfigs} we vary the properties of the main lens galaxy to ascertain which types of lens systems will provide stronger constraints on the Hubble constant.

Throughout this paper we assume a cosmology with $\Omega_M = 0.274$, $\Omega_\Lambda = 0.726$, and $H_0 = 71$ km s$^{-1}$ Mpc$^{-1}$.

\section{Methodology}
\label{sec:methods}

Our goal is to quantify ENV/LOS effects using realistic lensing beams. To examine ENV/LOS effects, we build mock beams using realistic matter distributions based on our photometric and spectroscopic observations. Our mass beams consist of two components: discrete mass structures such as group and galaxy halos, which are discussed in Section \ref{sec:massmodels}; and a smooth background mass density, which can account for under-dense regions (voids) and can be treated within our multi-plane lensing framework using the new approach described in Section \ref{sec:voids}. We ray trace from many source positions through these beams to produce simulated lensing observables including image positions, flux ratios, and time delays, as discussed in Section \ref{sec:observables}. Finally, we fit these lensing observables with different models (described in Section \ref{sec:lensmodels}) to understand how different modeling assumptions about the ENV/LOS lead to bias and/or scatter in the recovered values of lens galaxy parameters and the Hubble constant.

\subsection{Building Mass Beams from Observations \label{sec:massmodels}}
We begin by presenting our methodology for constructing beams from photometric and spectroscopic observations. Our beams are made of two components: a smooth background density and discrete objects, namely galaxies and group dark matter halos. We first consider the discrete structures in our beams.

\subsubsection{Discrete Mass Structures}
We include discrete mass structures using an updated version of the methodology in \citet{Wong11}. We start by introducing galaxies detected in our photometric observations into our beams. The photometric catalogs are based on 12 nights of imaging data from the 4m Mayall telescope on Kitt Peak and the 4m Blanco telescope on Cerro Tololo \citep{Williams06}. The photometric data are highly complete down to $I \leq 21.5$. For many of these objects, we have spectroscopic redshifts so we can include them at their measured 3D position (redshift and position on the sky). The spectroscopic sample is presented in \citet{Momcheva15}, and consists of 40 nights of 6.5m telescope observations using the LDSS-2, LDSS-3, IMACS, and Hectospec instruments (\citealt{Sluse16} present similar data for HE 0435$-$1223).
For LOS galaxies that lack spectroscopy, we must assign redshifts to get their 3D positions. If the field is in the SDSS footprint, we assign photometric redshifts from SDSS DR9 \citep{Ahn12, Csabai03}. If not, we draw from a smoothed version of the spectroscopic redshift distribution in that particular field. When considering the full error budget analysis (see Section \ref{sec:Environments} and K.\ C.\ Wong et al. 2017, in preparation), we marginalize over the unknown redshifts by drawing many realizations from the spectroscopic redshift distribution of the beam. 

Once we have the 3D position of the discrete mass structures (R.A., decl., and redshift), we must choose a mass distribution for the galaxy. We model each galaxy as a truncated singular isothermal sphere. This is a surprisingly good approximation for observed galaxy-scale lenses (e.g., \citealt{Rusin03, Gavazzi07, Koopmans09}, but see \citealt{Xu16}). Throughout our analysis, most of the galaxies included in our beams are significantly offset in projection from the main lens galaxy limiting our sensitivity to the exact mass distribution of the galaxies along the LOS.

The truncation radii are set to be equivalent to $r_{200}$. The radii and velocity dispersions are drawn from scaling relations derived from a galaxy--galaxy weak lensing analysis in the CFHTLS \citep{Brimioulle13}; we use luminosity scaling resolutions that evolve with redshift and include scatter. For $r_{200}$, \citet{Brimioulle13} give $r_{200}/r_0 = (L / L_0)^{\eta_r}$ where $L_0 = 1.6 \times 10^{10}\  h^{-2} L_\odot (1+z)$. For blue galaxies ($B-V \leq 0.7$), $r_0 = 133 \pm 3 \ h^{-1}$ kpc and $\eta_r = 0.40 \pm 0.08$, while for red galaxies ($B-V > 0.7$), $r_0 = 198 \pm 3 \  h^{-1}$ kpc and $\eta_r = 0.38 \pm 0.04$. For the velocity dispersion scaling relations, \citet{Brimioulle13} give $\sigma / \sigma_0 = (L / L0) ^{\eta_\sigma}$. $L_0$ is the same as for the radius scaling relation. For blue galaxies, $\sigma_0 = 123 \pm 3$ km s$^{-1}$ and $\eta_\sigma = 0.24 \pm 0.03$, while for red galaxies, $\sigma_0 = 173 \pm 2$ km s$^{-1}$ and $\eta_\sigma = 0.25 \pm 0.03$.

For lenses that are in a group or cluster, we treat the group member galaxies in a slightly different way. We compute a total dynamical mass for the group \citep{Girardi98,Momcheva06,Ammons14,Momcheva15, Wilson16} and apportion the mass between the group dark matter halo and the group galaxies. The dark matter halo is assumed to be a spherical NFW profile. The halo concentration is taken from the mass--concentration relation from simulations by \citet{Zhao09}, with an assumed scatter of 0.14 dex \citep{Bullock01}. The fraction of the total group mass assigned to the halo, $f_{\rm{halo}}$, is randomly drawn from a uniform distribution such that the group galaxies are not truncated below twice their effective radii, nor beyond their $R_{200}$ (the radius at which the mean density enclosed is 200 times the matter density at that redshift). The remainder of the mass, $(1-f_{\rm{halo}})M_{\rm{group}}$, is assigned to the group galaxies, which are scaled such that they have the same density at their truncation radii. All group galaxies and the group halo are assumed to be in the same redshift plane as the lens, even though there may be slight redshift differences due to peculiar velocities. 

Some lenses have groups along the LOS (e.g., B0712, \citealt{Fassnacht02}). Groups along the LOS of sight are treated in less detail. The galaxies are included as described above, but mass is not apportioned to a group NFW halo. For the current work, we are less concerned about reproducing exact lens fields and more concerned about producing \textit{realistic} matter distributions so that we can study how the LOS affects lens models. In future analyses we will account for groups along the LOS using an updated version of the methodology in \citealt{Ammons14}, see also \citealt{Wilson16}).

This methodology allows us to explicitly include matter that we see (e.g., galaxies). However, there may be matter that we do not observe like faint galaxies or dark matter filaments. We include the rest of that mass that we do not explicitly include in our discrete mass structures described above in a smooth, uniform mass component which is described in more detail in the next section.

\subsubsection{Setting the Smooth Background Density}

We now turn our attention to the underlying, smooth density component of our mass distributions. In traditional single-plane lensing, it is reasonable to assume that the lens galaxy does not significantly contribute to the overall geometry of the universe, and thus to place it on top of a smooth background density (the mean density of the universe).  As we begin to include more and more galaxies along the LOS, however, the cosmological effects of the mass included in our beams become non-negligible. Simply adding the galaxies on top of the mean density would lead to biased results. If our beams were 100\% complete, there should be no smooth background component at all.  In practice, we need to find an intermediate approach that scales the background density to account for incompleteness in our beams (which can arise, for example, from having flux-limited photometric and spectroscopic data).

There are several ways we could do this. One possibility is to start with an \emph{empty} beam and add both galaxies and a smooth density field.  However, we would still need to account for the mean density of the universe outside the beam when computing cosmic distances.  A different approach, which we prefer, is to start with a \emph{mean density} beam. We can then subtract a smooth density field that corresponds to the amount of mass that we have included with discrete mass structures (e.g., galaxies). The benefit is that we can then compute cosmic distances using the standard mean density universe expressions while still accounting for the lensing effects produced by having a smooth background density that is less than the mean density.

The key is that we need to estimate our ``mass completeness,'' which is defined to be the amount of mass that is explicitly included as discrete structures relative to the smooth mean density of the universe. We note that the mass completeness is not the same as the observational completeness (i.e., the fraction of galaxies that are successfully measured). We can use the mass completeness to subtract off the fraction of mass in our discrete structures from the mean density of the universe.

Consider a plane that represents the projection of a thin slab extending from proper distance $r_p$ to $r_p + dr_p$.  If the density contrast within this slab is $\drho_p = \rho_p - \bar\rho_p$, the lensing effects are characterized by a convergence
\begin{equation}
  d\kappa = \frac{\drho_p\,dr_p}{\Sigcrit}
\end{equation}
where the critical surface density for lensing is
\begin{equation}
  \Sigcrit = \frac{c^2}{4 \pi G} \frac{D_s}{D_\ell D_{\ell s}}
\end{equation}
Converting to comoving distances $X(z)$, we can write
\begin{equation}\label{eqn:dkappa}
  d\kappa = \frac{4 \pi G}{c^2}\frac{X (X_s - X)}{X_s} \frac{\drho_p}{(1+z)^2} dX.
\end{equation}
In order to achieve the desired subtraction, we set the density contrast to be
\begin{equation}
  \drho_p = -f(z) \bar{\rho}(z) 
\end{equation}
where $f(z)$ is the mass completeness in our models (which can vary with redshift), and the mean density of the universe at redshift $z$ is $\bar{\rho}(z) = \rho_{c} \Omega_m (1 + z)^3$, where $\rho_{c} = 3 H_0^2 / 8 \pi G$ is the critical density of the universe today.

\begin{figure}[t]
\begin{center}
\includegraphics[width=\columnwidth]{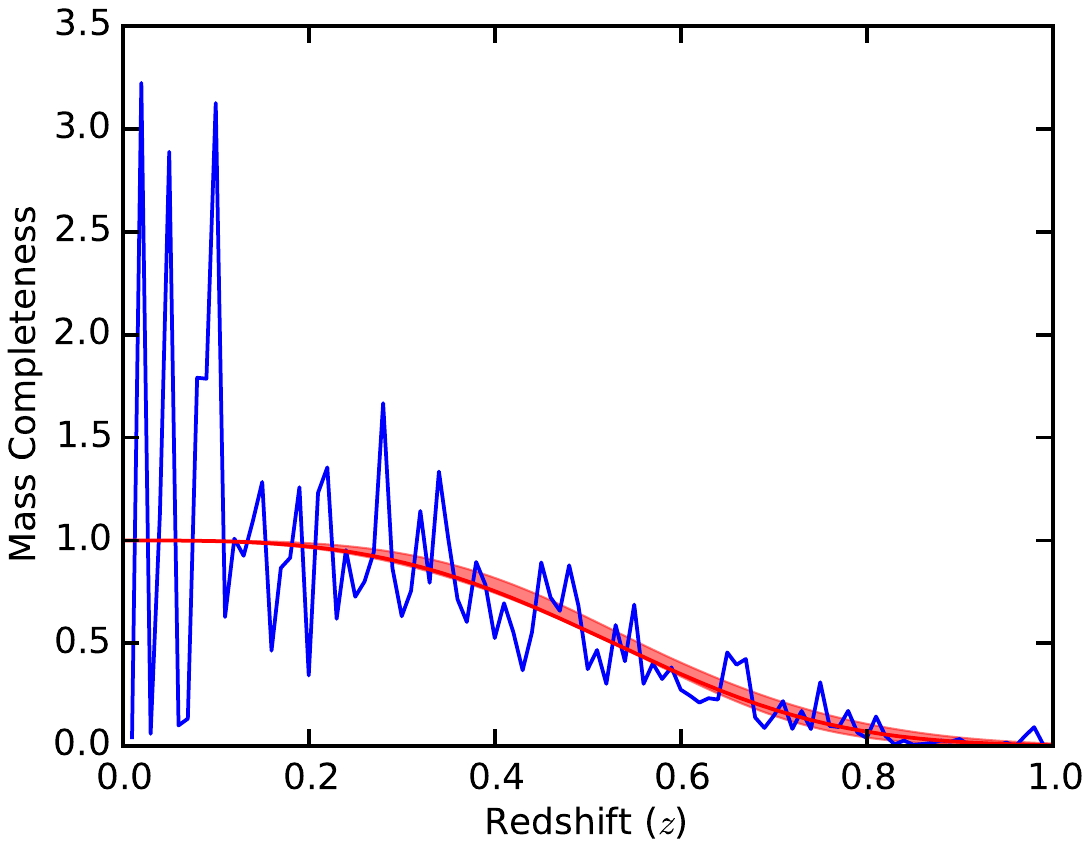}
\caption{\label{fig:mass_complete} Mass completeness as a function of redshift. The blue line shows the average mass per redshift bin, relative to a smooth universe, across the 23 fields for which we have constructed matter distributions. The red line shows the best fit to the data using a functional form of $f(z) = e^{-z^a / b}$ with best-fit parameters $a = 3.23$ and $b = 0.183$. The red band shows the 16--$84\%$ confidence interval derived from MCMC. We then use this mass completeness function to set the lensing contribution from the smooth background mass density. This allows us to self-consistently include voids. Because this is a fit to the average across beams, individual beams can have higher or lower than the mean density of the universe at a given redshift. The only assumption is that the average of our beams should be approximately the mean density of the universe.
}
\end{center}
\end{figure}

In practice, we parameterize the mass completeness, $f(z)$, with a simple analytic function. Using all 23 fields for which we have constructed mass distributions, we fit the average mass per redshift bin, relative to a smooth universe, using a function of the form $f(z) = e^{-z^a / b}$ as shown in Figure \ref{fig:mass_complete}. This function is a generalization of a Gaussian that captures the near-total completeness at low redshift (with $f(z=0) = 1$) and smoothly decreases in a similar manner to our observed mass completeness. We find best-fit values of $a = 3.23$ and $b = 0.183$. We then a Markov chain Monte Carlo (MCMC) model using the emcee package \citep{emcee} to estimate the uncertainties on the fitted parameters to the mass completeness. The confidence 1-$\sigma$ confidence interval is shown in the red band in Figure \ref{fig:mass_complete}. 

We adopt this mass completeness function for all subsequent analysis. Because we have used the average across all of our beams instead of fitting each beam individually, this allows for individual beams to be over- or underdense compared to the mean density of the universe. The only assumption is that the mean of our total sample is approximately the mean density of the universe in each redshift bin. However, \citet{Collett16} find that quad lenses in simulations tend to be in slightly overdense lines of sight ($\kappa_{\rm{ext}} = 0.009$). We can use this same methodology to estimate our completeness, but instead of correcting to the mean density, we would correct to the overdensity. We do not include this offset here as we attempting to create \textit{realistic} mass distributions, but this effect should be taken into account for analyses on real observations.

\subsubsection{Adding the Lens System to Our Beams}
The final ingredient for our matter distributions is the lens system itself. We treat the lensing potential of the main lens galaxy using a ellipsoidal power-law model given by
\begin{equation}
\label{eqn:powerlaw}
\kappa = \frac{b^{2-\eta}}{2 \xi^{2-\eta}}
\quad\mbox{where}\quad
\xi^2 = x^2 + \frac{y^2}{(1-e)^2}
\end{equation}

Here $\xi$ is the elliptical radius and $e$ is the ellipticity.  The power-law index $\eta$ is chosen so the enclosed mass scales as $M(r) \propto r^\eta$.
The Einstein radius for this model is $R_E = b \eta^{\eta - 2}$. Our initial models use $\eta = 1$ which corresponds to a singular isothermal ellipsoid (SIE).

We treat the azimuthal angle of the main lens galaxy and the position of the source as nuisance parameters to marginalize.  Specifically, we draw random values for the lens galaxy orientation from a uniform distribution.  

Now that we have the all the pieces of the mass distributions, we turn to the machinery we need to use these mass distributions to generate mock lensing observables.

\subsection{Generating Mock Lensing Observables} \label{sec:observables}

\subsubsection{Multi-plane Lensing}
As we are interested in how mass in the ENV/LOS affects lensing measurements, we must go beyond the standard lens equation that only includes a single lens. Instead we use the multi-plane lens equation (which makes the thin-lens approximation at each redshift of interest; \citealt{PLW}, see \citealt{Petkova14} for an implementation of multi-plane lensing in the GLAMER code, \citealt{Metcalf14}) that traces a light ray from the observer to the source, bending the ray at each redshift plane of interest. The full multi-plane lens equation is given by:
\begin{equation}
\x_{j} = \x_1 - \sum_{i=1}^{j - 1} \beta_{i\,j} \al_i(\x_i),
\label{eqn:full_mp}
\end{equation}
where $\al_i$ is the deflection in plane $i$ and
\begin{equation}
\beta_{i\,j} = \frac{D_{i\,j} D_s}{D_{j}D_{i\,s}}.
\end{equation}
The Jacobian matrix for the mapping between the coordinates on the sky and the coordinates in plane $j$ is
\begin{equation} \label{eq:multi-A}
  \A_{j} \ = \ \I - \sum_{i=1}^{j-1} \beta_{ij} \frac{\partial\al_i}{\partial\x_i}
       \frac{\partial\x_i}{\partial\x_1}
  \ =\ \I - \sum\limits_{i=1}^{j-1} \beta_{ij} \GammaMat_i \A_i\,,
\end{equation}
where $\GammaMat_i$ is the ``tidal tensor'' in plane $i$ and is given by
\begin{equation}
  \GammaMat_i = \frac{\partial\al_i}{\partial\x_i}
  = \left[\begin{array}{cc}
    \kappa_i + \gamma_{\mathrm{c},i} & \gamma_{\mathrm{s},i} \\
    \gamma_{\mathrm{s},i} & \kappa_i - \gamma_{\mathrm{c},i}
  \end{array}\right].
\end{equation}
We have defined the convergence ($\kappa$) and shear ($\gamma$) components from galaxy $i$ in terms of the lensing potential, $\phi_i$ (defined as $\al_i(\x_i) = \nabla\phi_i(\x_i)$), to be
\begin{eqnarray}
  \kappa_i &=& \frac{1}{2} \left( \frac{\partial^2\phi_i}{\partial x_i^2} +
    \frac{\partial^2\phi_i}{\partial y_i^2} \right) , \\
  \gamma_{\mathrm{c},i} &=& \frac{1}{2} \left( \frac{\partial^2\phi_i}{\partial x_i^2} -
    \frac{\partial^2\phi_i}{\partial y_i^2} \right) , \\
  \gamma_{\mathrm{s},i} &=& \frac{\partial^2\phi_i}{\partial x_i \partial y_i}\ .
\end{eqnarray}
The magnification is the inverse of the Jacobian matrix for the source plane, $\mu = \A_s^{-1}$.
The time delays are given by
\begin{equation}
  T = \sum_{i=1}^{s-1}  \frac{1}{2} \tau_{i\,i+1} \left |\x_{i+1} - \x_i \right |^2 - \tau_{i\,s} \phi_i(\x_i),
\label{eqn:full_t}
\end{equation}
where 
\begin{equation}
\tau_{ij} =\frac{1+z_i}{c} \frac{D_i D_j}{D_{ij}}
\end{equation}.

Using these equations to generate lensing observables from discrete mass structures is straightforward, but including the smooth density for voids requires more care, which we will now discuss.

\subsubsection{Multi-plane Lensing with Voids}\label{sec:voids}
 We need the three-dimensional information (sky position and redshift) inherent in the multi-plane lensing formalism, making it attractive to study lensing effects due to the ENV/LOS. However, the multi-plane framework is intrinsically built to handle discrete structures, so we now must generalize the multi-plane formalism to include a smooth background density. In our case, we will subtract from the mean density (so that we do not double count the mass in our models) by inserting redshift planes with negative mass density (i.e., negative convergence) in between the planes that contain galaxies. In the limit of a large number of thin planes, the sums over ``void'' planes becomes integrals that we now derive.

Suppose there are only void planes between $z_i$ and $z_j$, with $\GammaMat_k \rightarrow \Delta \kappa_k \I$ for all $i < k < j$. If we consider the lens equation (e.q. \ref{eqn:full_mp}) for void plane $i+1$, we see that
\begin{equation}
\x_{i+1} = \x_1 - \sum\limits_{k=1}^{i}\beta_{k\, i+1} \al_k(\x_k).
\end{equation}
Then stepping through to $i+2$, we find that
\begin{multline}
\x_{i+2} = (1 - \beta_{i+1 \, i+2} \Delta \kappa_{i+1}) \x_1 \\ 
-  \sum\limits_{k=1}^{i}(\beta_{k\, i+2} -\beta_{i+1 \, i+2} \beta_{k\, i+1} \Delta \kappa_{i+1}) \al_k. 
\end{multline}
This generalizes to 
\begin{equation}
\x_j = \betahat_{0 j} \x_1 - \sum \limits^{j - 1}_{i = 1} \betahat_{i j} \al_i (\x_i)
\end{equation} where we have defined the scaled set of distance ratios to be
\begin{equation}
\label{eqn:betahatdef}
\betahat_{i j}  = \beta_{i j} - \sum \limits_{k = i +1}^{j -1} \betahat_{i k} \beta_{k j} \Delta \kappa_k,
\end{equation}
and used the identity that if we take the observer to be in plane $i = 0$, $\beta_{0 j} = 1$. The sum in equation \ref{eqn:betahatdef} is now only over non-convergence planes.

The Jacobian matrix (eq.\ \ref{eq:multi-A}) generalizes to
\begin{equation}
 \label{eqn:full_mp_voids} \\
\A_j = \betahat_{0 j} \I - \sum \limits^{j -1}_{i = 1} \betahat_{i j} \GammaMat_i \A_i.
\end{equation}

In the limit of a large number of thin void planes, the sum in equation \ref{eqn:betahatdef} becomes a continuous integral:
\begin{equation}
\label{eqn:betahatint}
\betahat_{i j} = \beta_{ i j} - \int_{z_i}^{z_j} \betahat(z_i, z) \beta(z, z_j) d\kappa(z)
\end{equation}
This expression is recursive and thus unwieldy in practice, but it suggests that $\betahat$ can be described by a differential equation. We are going to take derivatives of \ref{eqn:betahatint} with respect to comoving distances, so it is first useful to examine derivatives of $\beta_{i j}$. In terms or comoving distances, $X_i$, 
\begin{eqnarray}
\beta_{i j} &=&\frac{D_{i j} D_s}{D_{i s} D_j} = \frac{(X_j - X_i) X_s}{X_j (X_s - X_i)} \\
\frac{d \beta_{i j}}{d X_j} &=& \frac{X_i X_s}{X_j^2 (X_s - X_i)}\\
\label{eqn:d2betahatdbetahat} 
\frac{d^2 \beta_{i j}}{d X_j} &=& -2 \frac{X_i X_s}{X_j^3 (X_s - X_i)} = -\frac{2}{X_j} \frac{d \beta_{i j}}{d X_j} 
\end{eqnarray}

Now, differentiating equation \ref{eqn:betahatint} with respect to $X_j = X(z_j)$ yields
\begin{equation}
\label{eqn:dbetahat}
\frac{d \betahat_{i j}}{d X_j} = \frac{d\beta_{i j}}{d X_j} - \int_{X_i}^{X_j} \betahat(z_i, z) \frac{d \beta(z, z_j)}{d X_j} \frac{d \kappa}{d X} d X. 
\end{equation}
where we use $\beta(z_j, z_j) = 0$.  The second derivative is
\begin{multline}
\label{eqn:d2betahat}
\frac{d^2 \betahat_{i j}}{d X_j^2} = \frac{d^2 \beta_{i j}}{d X_j^2} -  \betahat_{i j} \frac{d \kappa}{d X_j} \left. \frac{d \beta(z, z_j)}{d X_j}  \right|_{z \rightarrow z_j} \\
 - \int_{X_i}^{X_j} \betahat(z_i, z) \frac{d^2 \beta(z, z_j)}{d X_j^2} \frac{d \kappa}{d X} d X.
\end{multline}
Now we combine equations \ref{eqn:dbetahat} and \ref{eqn:d2betahat} in the following way:
\begin{multline}
\frac{d^2 \betahat_{i j}}{d X_j^2} + \frac{2}{X_j} \frac{d \betahat_{i j}}{d X_j} =  \\
\left[\frac{d^2 \beta_{i j}}{d X_j^2}  + \frac{2}{X_j} \frac{d \beta_{i j}}{d X_j}\right]
- \betahat_{i j} \frac{d \kappa}{d X_j} \left. \frac{d \beta(z, z_j)}{d X_j}  \right|_{z \rightarrow z_j} \\ 
 - \int_{X_i}^{X_j} \betahat(z_i, z)\left[ \frac{d^2 \beta(z, z_j)}{d X_j^2}  + \frac{2}{X_j} \frac{d \beta_{i j}}{dX_j}\right] \frac{d \kappa}{d X} d X.
\label{eqn:betahatcombine}
\end{multline}
The terms in square brackets vanish by equation \ref{eqn:d2betahatdbetahat}. Then using equations \ref{eqn:dkappa} and \ref{eqn:dbetahat} in the remaining term on the right-hand side yields
\begin{equation}
\frac{d^2 \betahat_{i j}}{d X_j^2} + \frac{2}{X_j}\frac{d \betahat_{i j}}{d X_j} = - \frac{4 \pi G}{c^2} \frac{\drho_p}{(1 + z_j)^2} \betahat_{i j}.
\end{equation}
This is our desired differential equation, which can be solved using standard numerical techniques.  Since it is a second-order equation, we need two boundary conditions. The first is $\betahat_{i i} = 0$. The second condition is based on the first derivative with 
\begin{equation}
\left. \frac{d \betahat_{i j}}{d X_j} \right|_{j \rightarrow i} = \frac{D_s}{D_i D_{i s}}.
\end{equation}

We can follow a similar set of arguments for time delays. As before, if there are only void planes between $z_i$ and $z_j$, following a similar analysis to Section 4 in M14, we find that the generalized multi-plane expression for the time delay (eq.\ \ref{eqn:full_t}) is 
\begin{multline}
T = \frac{1}{2} \tauhat_{0 s} \x_s \cdot \x_1 + \\ \sum \limits_{i = 1}^{s-1} \left[ \frac{1}{2} \tau_{\ell s} \x_\ell \cdot \al_\ell - \frac{1}{2} \tauhat_{\ell s} \x_s \cdot \al_\ell - \tau_{\ell s} \phi_{\ell}(\x_\ell)\right],
\end{multline}
where the sum is again over non-convergence planes.  We emphasize that the terms with potentials ($\phi_i$) and $\x_\ell \cdot \al_\ell$ have a coefficient of $\tau_{\ell s}$, \emph{not} $\tauhat_{\ell s}$; the terms with $\x_s$ include light propagation effects so they include $\tauhat$, while the other terms only depend on the indivdual lens plane so they only have a regular $\tau$.

The generalized time scales follow the relation
\begin{equation}
\tauhat_{i j} = \tau_{i s} - \sum\limits_{k = i + 1}^{j - 1}\betahat_{i k} \tau_{k s} \Delta \kappa_k.
\end{equation}
Defining again the $i=0$ plane to be the observer, we find that $\tau_{0 j}=0$, though we note that $\tauhat_{0\, j}$ may be nonzero. 
The continuous limit is
\begin{equation}
\tauhat_{i j}  = \tau_{i s} - \int_{X_i}^{X_j} \betahat(z_i, z) \tau(z, z_s) \frac{d \kappa}{d X} dX
\end{equation}
We now take a derivative of $\tauhat_{i j}$ with respect to $X_j$ and obtain
\begin{equation}
\frac{d \tauhat_{i j}}{d X_j}  = -\betahat_{i j} \tau_{j s} \frac{d \kappa}{d X_j}. 
\end{equation}
This is a first-order differential equation, so we only need one boundary condition. The simplest choice is $\tauhat_{i j} \rightarrow \tau_{i s}$ as $j \rightarrow i$ because in that case there is no void correction.  

We now use this multi-plane lensing formalism (including voids) to ray trace through our matter distributions to generate our lensing observables. 

\subsubsection{Lensing observables}
We generate mock lensing observables by ray tracing from a mock ``source`` at some position and redshift. All of our simulations and modeling use the \emph{lensmodel} package \citep{Keeton01}.  We generate the following mock lensing observables by ray tracing from a mock ``source`` at some position and redshift using the full multi-plane lens equation (eq.\ \ref{eqn:full_mp_voids}, including the correction for voids described above): image configuration, image positions, time delays, and flux ratios (similar to \citealt{Petkova14}).

To summarize, two-image lens systems have the following observables:
\begin{itemize}
\item $2\times 2$ image  positions ($x,y$)
\item 2 fluxes
\item 1 time delay 
\item $1\times 2$ lens galaxy position ($x,y$)
\end{itemize}
giving a total of nine constraints.

By contrast, four-image systems have:
\begin{itemize}
\item $4\times 2$ image positions ($x,y$)
\item 4 fluxes
\item 3 time delays
\item $1\times 2$ lens galaxy position ($x,y$) 
\end{itemize}
giving a total of 17 constraints.

We imagine observing these lenses with fiducial measurement uncertainties of 3 mas for the positions of the lensed images and the main galaxy, $5\%$ for the fluxes, and 1 day for the time delays. These values are typical of observations with instruments such as \emph{HST} and monitoring campaigns such as COSMOGRAIL \citep{Eigenbrod05}. We do not explicitly add observational scatter to the mock data; preliminary tests indicate that such measurement noise introduces a ``floor'' in the $\chi^2$ values and the scatter in recovered lens parameters but does not qualitatively change our results.

We choose source positions from a magnification weighted distribution to approximate the magnification bias that is present in real observations \citep[see][]{Keeton04}. Overall, we generate a sample of 300 quad and 300 double mock image configurations for each mass beam.

\subsection{Fitting Mock Lensing Observables  \label{sec:lensmodels}}

We can then fit the simulated observables as we would for real lens systems. We consider models that treat the ENV/LOS in progressivly more realistic ways to see if we can recover the true, input lens galaxy parameters and the Hubble constant. 

The models all have the following 10 free parameters for the main lens system:
\begin{itemize}
\item Source position ($x,y$)
\item Source flux
\item Hubble constant ($h$)
\item Einstein radius of the main lens galaxy
\item Lens galaxy position ($x,y$).
\item Lens galaxy ellipticity and orientation ($e,\theta_e$)
\item Lens galaxy power law index ($\eta$)
\end{itemize}

In all cases, the model parameters are varied to find the best fit and allowed range.  We seek to understand how the assumptions and approximations in the various models cause parameters to shift away from their true values. 

For the recovered lens quantities, we use MCMC methods to calculate the full posterior probability distribution. When presenting marginalized results for a given lens parameter, we show the median value of the fitted parameters and estimate the scatter by measuring half of the difference between the $16^{th}$ and $84^{th}$ percentiles. When plotting results, we show fractional changes in the fitted parameters to emphasize the variation.

In some cases it may be reasonable to put priors on the lens galaxy ellipticity and orientation based on the light profile \citep[e.g.,][]{Bolton08}, but in lensing there is not always a strong relation between the light and mass \citep[][and references therein]{Bruderer16}. Therefore we do not place any priors on the lens galaxy parameters.  We do assume a weak Gaussian prior on the Hubble constant, $h = 0.71 \pm 0.3$, which ensures that models are reasonable but otherwise has little effect on the results for quad lenses.  The prior plays a stronger role in double lenses since they are under-constrained (see below).  Overall, the results from double lenses are qualitatively similar to those from quad lenses but considerably broader. For the remainder of the discussion we focus on quad lenses because they are better constrained.

The details of the individual models we consider here are in the following subsections.

\subsubsection{Lens-Only}\label{sec:lensonly}
The ``Lens-Only'' model ignores ENV/LOS effects altogether and is therefore the simplest model. This model fits the following form of the lens equation:
\begin{equation}
\x_s = \x_1 - \tilde{\al_\ell}(\x_1),
\end{equation}
where the tilde indicates that the parameters in the lens potential are allowed to vary.
Given the numbers of parameters and constraints, quad lenses are well constrained with 7 degrees of freedom (dof). Double lenses are, however, under constrained with $-1$ dof.

The Lens-Only model ignores the ENV/LOS entirely, so it serves as a control test: the best-fit values for the lens galaxy parameters and the Hubble constant move away from the truth in an attempt to account for the LOS/ENV effects, allowing us to characterize the overall importance of ENV/LOS effects.

\subsubsection{Lens+Shear}\label{sec:lens+shear}
The ``Lens+Shear'' model attempts to account for ENV/LOS effects by fitting an external shear in the main lens plane \citep[e.g.,][]{Suyu13}. The Lens+Shear model has two additional free parameters that characterize the external shear; we use the pseudo-Cartesian components of the shear, $\gamma_c = \gamma \cos 2\theta_\gamma$ and $\gamma_s = \gamma \sin 2\theta_\gamma$.
The lens equation for the Lens+Shear models takes the following form:
\begin{equation}
\x_s = (\I - \tilde{\GammaMat}) \x_1 - \tilde{\al_\ell}(\x_1),
\end{equation}
where the tilde again indicates what is allowed to vary in the fits; in this case the main lens galaxy potential and the external shear.
For the Lens+Shear models, quad lenses have 5 dof and double lenses are underconstrained with $-3$ dof.
Because the Lens+Shear model has a different number of dof, care must be taken when comparing the results to the other models. We scale the $\chi^2$ value by the 95\% confidence limit for the number of dof to ensure a fair comparison across model types.

The Lens+Shear model ignores any 3D structure of the LOS. Comparing the Lens+Shear model to more realistic (and therefore complex) models allows us to quantify the importance of redshift effects in modeling realistic lens systems.

\subsubsection{Lens+3D Tides}\label{sec:3Dtides}
The ``Lens+3D Tides" model uses our multi-plane framework developed in M14 (see also \citealt{Schneider97}). This includes the nonlinear effects due to 3D LOS structure, but ignores higher-order terms beyond the tidal approximation.

The Lens+3D Tides models replace the traditional tidal tensor with 3D versions. The expressions for the lens equation, magnification tensor, and time delay are as follows:
\begin{eqnarray}
\x_{s} &=& \B_s \x_1 - \C_{\ell s} \tilde{\al}_\ell(\B_\ell \x_1),\label{eqn:lenseqn_3Dtide}\\
\A_{s} &=& \B_s - \C_{\ell s} \frac{\partial \tilde{\al}_\ell}{\partial \x_\ell} \A_\ell,\\
T &=& \frac{1}{2} \x_s \cdot \F_s \x_1 + \frac{1}{2} \tau_{\ell s} \B_\ell \x_1 \cdot \tilde{\al}_\ell(\B_\ell \x_1) \nonumber \\
&&-\frac{1}{2}\x_s \cdot \G_{\ell s}\tilde{\al}_\ell(\B_\ell \x_1) -\tau_{\ell s}\tilde{\phi}_\ell(\B_\ell \x_1),
\end{eqnarray}
where $\ell$ is the index of the redshift plane of the main lens galaxy. The tildes denote that we vary the main lens potential, using the same notation as above.
All of the tidal effects are characterized by the following ``3D Tidal Tensors'':
\begin{eqnarray}
\B_j &=& \betahat_{0 j} \I - \sum\limits_{i=1,i \neq \ell}^{j-1}\betahat_{i j}\GammaMat_i \B_i, \\
\C_{\ell j} &=& \betahat_{\ell j} \I - \sum\limits_{i=\ell+1, i\neq \ell}^{j-1} \betahat_{\ell i}\GammaMat_i\C_{\ell i}, \\
\F_j &\equiv& 
  \tauhat_{0 s} \I - \sum\limits_{i=1,i\neq \ell}^{j-1} \tauhat_{is}\GammaMat_i\B_i, \\
\G_{\ell j}&\equiv& 
 \tauhat_{\ell j}\I- \sum\limits_{i=\ell+1,i \neq \ell}^{j-1} \tauhat_{is}\GammaMat_i\C_{\ell i}.
\end{eqnarray}
Like the Lens-Only models, we only allow the main lens galaxy parameters to vary, giving 7 dof for quad lenses and $-1$ dof for doubles.

We have written the above expressions for the ``3D Tidal Tensors'' in terms of the $\betahat$ and $\tauhat$ parameters that we derived above. This is a generalization from M14 to include the effects of voids in these tidal tensors (this can be shown using similar arguments as in Section \ref{sec:voids}). In this case, the light propagation happens through the 3D Tidal tensors; as such $\betahat$ and $\tauhat$ appear in the definition of the tensors, but $\tau$ appears in the time delay expression for terms that only depend on the main lens plane.

These ``3D Tidal Tensors'' are not allowed to vary in this analysis (though in principle they could be) so they can be computed once at the start of any lens modeling analysis and stored for repeated use yielding a large increase in computational efficiency. We also note that the 3D Tidal Tensors depend on ratios of cosmic distances, so these matrices are only constant if the cosmological parameters besides $H_0$ are fixed. When other cosmological parameters are allowed to vary, e.g., constraining $\Omega_{\Lambda}$, it is necessary recompute these matrices. One possibility to overcome this computational challenge is to compute a grid of 3D Tidal Tensors and interpolate to a given set of cosmological parameters.

The lens equation in eq.\ \ref{eqn:lenseqn_3Dtide} looks similar to the lens equation in the Lens+Shear model, but with two key differences. The first is that 3D redshift effects are explict in the $\B_s$ tensor, instead of defining an effective shear in the lens plane. The second, more important, difference is that the $\B_\ell$ tensor appears inside the argument of the deflection of the main galaxy. This leads to nonlinear effects that cannot be generally rewritten in terms of an effective shear. By comparing the results from Lens+Shear and Lens+3D Tides models, we can explicity distinguish these nonlinear effects and assess their importance for cosmology. 

\subsubsection{3D Lens}\label{sec:3Dlens}
The final type of model we consider is the most general. These models include the most important galaxies (ranked according to a criterion discussed in Section \ref{sec:Individual}) including higher-order terms (not using the tidal approximation) while the rest of the ENV/LOS galaxies are included in the 3D Tidal Tensors similar to the Lens+3D~Tides model. We have called these the ``3D Lens'' models.

 Our lens models typically contain hundreds of components, but we expect that many of the perturbing galaxies do not significantly contribute to the lens equation beyond their shear and convergence. In our previous paper (M14; see \citealt{Schneider14} for an alternative derivation), we presented a hybrid framework for multi-plane lensing that incorporates all of the tidal planes (weak lenses) into the ``3D Tidal Tensors'' so the sum in the multi-plane lens equation (eq.\ \ref{eqn:full_mp}) includes only ``exact'' lens planes (strong lenses, i.e., those not treated with the tidal approximation). Our framework recovers the full multi-plane lens equation when all of the redshift planes are treated as ``exact'' planes. Even when it uses the tidal approximation, our framework still includes nonlinear effects associated with having mass at different redshifts. 
 
 The primary difference in the lens equation between the Lens+3D~Tides and the full 3D lens model is that the $\ell$ becomes a free index that we use to sum over ``exact'' planes. We then realize that the $\B$ and $\C$ matrices are related, as are the $\F$ and $\G$ matrices:
\begin{equation}\label{eqn:BC-FG}
\B_j = \C_{0 \, j}
\qquad\mbox{and}\qquad
\F_{j} = \G_{0 \, j}.
\end{equation}

For an arbitrary mix of strong lenses (``exact planes'') and weak lenses (whose effects are captured in the 3D Tidal Tensors), the 3D Tidal Tensors become
\begin{eqnarray}
\C_{\ell j} &=& \betahat_{\ell j} \I - \sum\limits_{i=\ell+1, i\not\in\{\ell_\mu\}}^{j-1} \betahat_{\ell i}\GammaMat_i\C_{\ell i},\\
\G_{\ell j}&\equiv& 
 \tauhat_{\ell j}\I- \sum\limits_{i=\ell+1,i\not\in\{\ell_\mu\}}^{j-1} \tauhat_{is}\GammaMat_i\C_{\ell i}.
\end{eqnarray}

 The generalized expressions for the lens equation, magnification tensor, and time delay are as follows:
\begin{eqnarray}
\x_{j} &=& \C_{0\, j} \x_1 - \C_{\ell j} \tilde{\al}_\ell(\x_\ell) - \sum_{i \in \{\ell_\mu<j\}} \C_{i j} \al_i(\x_i),\label{eqn:lenseqn}\\
\A_{j} &=& \C_{0 \, j} - \C_{\ell j} \frac{\partial \tilde{\al}_\ell}{\partial \x_\ell} \A_\ell - \sum_{i \in \{\ell_\mu<j\}} \C_{i j} \GammaMat_i \A_i,\\
T &=& \frac{1}{2} \x_s \cdot \G_{0 \,s} \x_1 + \frac{1}{2}\tau_{\ell s} \x_\ell \cdot \tilde{\al}_\ell -\frac{1}{2}\x_s \cdot \G_{\ell s}\tilde{\al}_\ell -\tau_{\ell s}\tilde{\phi}_\ell \nonumber \\
&&+ \sum\limits_{i \in \{\ell_\mu\}}\left[\frac{1}{2}\tau_{i s} \x_i \cdot \al_i -\frac{1}{2}\x_s \cdot \G_{i s}\al_i -\tau_{i s}\phi_i \right],
\end{eqnarray}
where the sums run over ``exact'' planes (which do not use the tidal approximation) and $\al_\ell$ is the deflection of the main lens. Again the tildes denote the quantities we allow to vary in our fits. While several galaxies may be included as ``exact'' planes, only the main lens galaxy is allowed to vary. Like the Lens-Only and Lens+3D Tides models, quad lenses have 7 dof and double lenses are under-constrained with $-1$ dof. 

Note in eq.\ \ref{eqn:lenseqn} that each ``exact'' plane needs to be evaluated using the positions $\x_i$, which are not generally the same as the positions $\x_1$ on the sky and must be computed with the lens equation. This distinction gives rise to nonlinearities that cannot be mimicked by an effective shear, in the same way as was discussed for the Lens+3D Tides model in Section \ref{sec:3Dtides}. Our framework is therefore more accurate than fitting a simple shear in the lens plane, but it is also more efficient than the full multi-plane lens equation because the recursive sums include only ``exact'' planes.  It fills the gap between using the full multi-plane lens equation (which can be computationally expensive) and treating everything as a simple external shear (which omits higher-order effects that can be significant for objects projected near the lens and does not include nonlinear redshift effects).

 By comparing the results from the Lens+3D~Tides models to the results from the full 3D Lens models, we isolate the effects from higher-order terms from the nonlinear redshift effects. Also, as we include more and more galaxies exactly (without using the tidal approximation) the 3D lens model becomes identical to the lens equation that was used to generate the mock lensing observables. This allows us to probe the fundamental limitations of recovering lens galaxy parameters and cosmological information from lens modeling analyses.  

 We now have all of the machinery in place to study the ENV/LOS contribution to lensing. We now turn to the results of our simulations.

\section{Results}
\label{sec:results}
To understand the contribution of the ENV/LOS to lensing, we begin by considering a toy system, but then progressively add complexity/realism to our simulations. In Section \ref{sec:Individual}, we perform the end-to-end analysis described in Section \ref{sec:methods}, but using a simplified mass distribution that consists of a main lens galaxy and a single perturbing galaxy. This allows us to build intuition about the effects due to individual ENV/LOS galaxies and to quantify a perturber's importance. With these lessons in hand, we expand our analysis in Section \ref{sec:Environments} by building and characterizing realistic mass distributions that contain hundreds of ENV/LOS galaxies (following Section \ref{sec:massmodels}). In Section \ref{sec:fitting}, we use these realistic mass distributions to generate mock lensing observables (see Section \ref{sec:observables}) which we then fit (see Section \ref{sec:lensmodels}) to determine how well different models recover the input galaxy parameters and cosmology. Finally, in Section \ref{sec:ImageConfigs}, we place different lens systems into the ENV/LOS mass distributions to understand how the properties of the main lens system alter its sensitivity to ENV/LOS effects when inferring the Hubble constant.

\subsection{Which Individual Environment/LOS Galaxies Are the Most Important?}
\label{sec:Individual}

While there are hundreds of galaxies in our beams, they are not equally important for lensing.  Conceptually, we want to understand what properties control an individual perturber's influence on the lensed images; we aim to specifically test the effects of mass, projected offset from the main lens galaxy, and redshift of a perturbing galaxy. To this end, we begin by considering our full analysis, but using a simplified mass distribution that consists of the main lens galaxy and one perturbing galaxy.

We initially control for redshift, placing the perturbing galaxy at the same redshift as the main lens but varying the mass and projected offset (Section \ref{sec:toylos}). We then move the perturbing galaxy in redshift (Section \ref{sec:frontback}). We use both equations and simulations to identify a quantity, based on the flexion produced by a perturbing galaxy, that characterizes how much a given galaxy affects the recovered value of the Hubble constant. Based on the results from M14, we also expect there to be a difference between perturbers in the foreground and background of the main lens due to nonlinear effects that we aim to quantify. 

\subsubsection{Massive Galaxies Projected Near the Lens are More Important \label{sec:toylos}}

We begin our characterization of individual perturbing galaxies with the simplest case: we examine the main lens galaxy and  perturbing galaxy in the same plane (removing any redshift effects). We first derive analytic expectations for how an individual perturbing galaxy affects the lens potential, and then, using simulations, show that the deviations in the recovered Hubble constant scale in the same way as the lens potential perturbations. We note that to derive our expectations for the behavior of our models, we use some simplifying assumptions (e.g., a point mass perturber), but we that we only make these assumptions when deriving our criterion to characterize the individual contribution of a perturbing galaxy along the LOS, but otherwise nowhere else in the fits.\footnote{When fitting our realistic lens beams, we use truncated SIS potentials to model the galaxies in the ENV and along the LOS (see Section \ref{sec:massmodels}).}

Consider a single main lens galaxy with deflection, $\al_{g}$, and a single perturbing galaxy with deflection, $\al_p$. Both are at the same redshift in this case.

If the perturbing galaxy does not overlap the lensed images, its gravitational effect is the same as a point mass so we can write the lens potential as
\begin{equation}
\phi_p = R_p^2 \ln \left| \x - \vec{r} \right| 
\end{equation}
where $R_p$ is the Einstein radius of the perturber, $\x$ is the image position in the redshift plane of the perturber, and $\vec{r}$ is the position of the perturber. If we let $|\x| = x$, $|\vec{r}| = r$, and $\theta=$ the angle between the perturber and the image position as measured from the origin, we can rewrite the potential using the law of cosines as
\begin{equation}
\phi_p = \frac{1}{2}R_p^2 \ln(r^2 + x^2 - r x \cos\theta).
\end{equation}
If the perturbing galaxy is far from the lensed images, then $x \ll r$ and we can expand the potential in a Taylor series:
\begin{eqnarray}
\phi_p &=& R_p^2 \left[ \ln(r) - \cos(\theta) \frac{x}{r}  - \frac{1}{2} \cos(2\theta) \frac{x^2}{r^2} \right. \nonumber\\
&& \qquad\left. - \frac{1}{3}\cos(3\theta)\frac{x^3}{r^3}  + \ldots\right].
\end{eqnarray}
This expression can be generalized beyond a point mass model as
\begin{equation}
\phi_p = \phi(0) + \alpha^i(0) x^i + \frac{1}{2}\Gamma^{ij} x^i x^j+ \frac{1}{6} \sF^{ijk} x^i x^j x^k + \ldots 
\end{equation}
where we have adopted the Einstein convention of summing over repeated indices and defined the tidal and flexion\footnote{Throughout this work, we refer to  third-order terms collectively as ``flexion'' terms, but see \citet{Bacon06} for more discussion.} tensors:
\begin{eqnarray}
\Gamma^{ij} &\equiv& \left. \frac{\partial^2 \phi_p}{\partial x^i \partial x^j} \right|_{x=0} , \\
\sF^{ijk} &\equiv&  \left. \frac{\partial^3 \phi_p}{\partial x^i \partial x^j \partial x^k} \right|_{x=0} .
\end{eqnarray}
For a point mass, the tidal and flexion amplitudes scale as $\Gamma \propto R_p^2/r^2$ and $\sF \propto R_p^2/r^3$.

\begin{figure}[t]
\begin{center}
\includegraphics[width=\columnwidth]{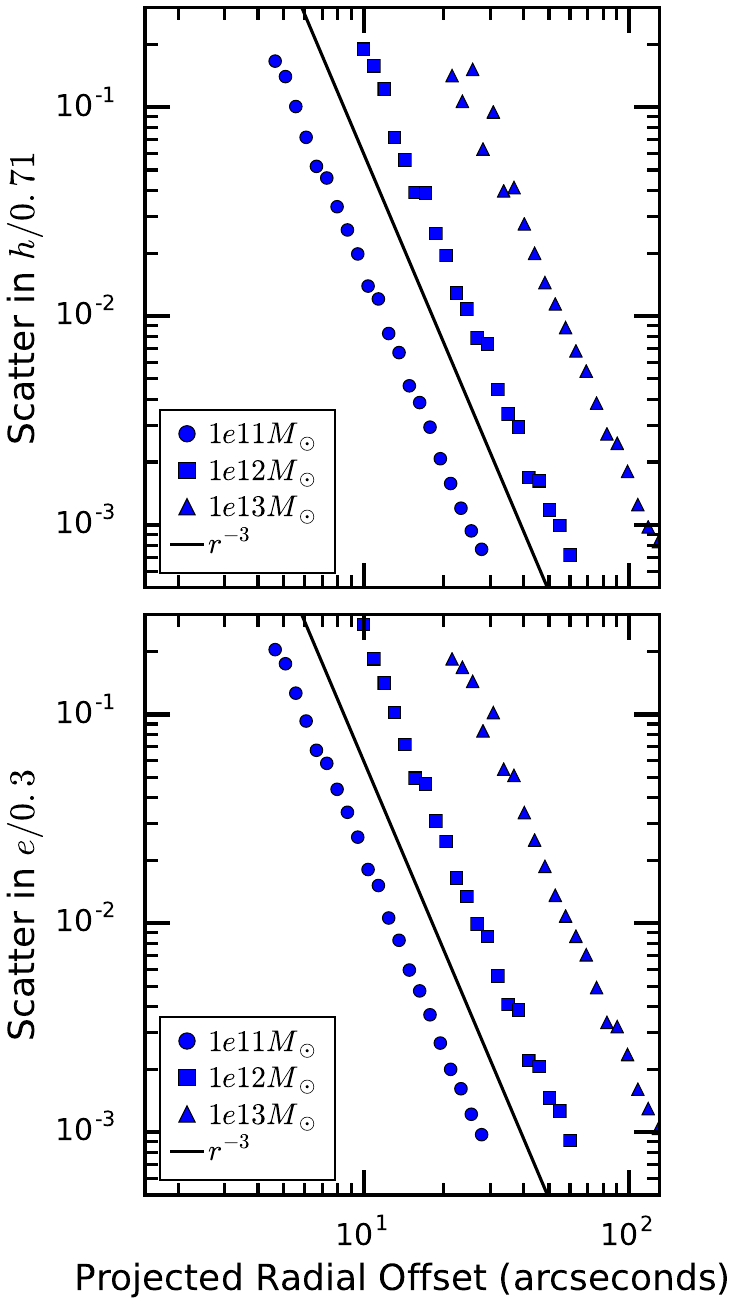}
\caption{\label{fig:toyr3} Deviations in lens model parameters for simulations with a single perturbing galaxy at the same redshift as the main lens galaxy.  The horizontal axis is the radial offset of the perturber, and the vertical axis is the scatter in recovered values of the Hubble constant ($h$, top) and ellipticity ($e$, bottom) recovered from Lens+Shear models. The points show results from simulations for three decades in the perturber's mass. The parameter deviations follow an $r^{-3}$ power law (indicated with a line) because Lens+Shear models omit third-order terms and higher in the lens potential. The deviations also scale with perturber mass as $M \propto R_p^2$. Thus, understanding lens potential perturbations provides a robust way to characterize a perturbing galaxy's effect on the Hubble constant.%
}
\end{center}
\end{figure}

In the expansions, the zeroth-order terms correspond to shifts in the zeropoint of the potential, which are unobservable.  The first-order terms correspond to a uniform deflection, which can be absorbed by translating the source plane.  Therefore only the terms at second order and higher have observable consequences.

To quantify these environmental effects, we employ a simplified mass distribution with an SIE main lens galaxy and a point mass perturber.  We choose the Einstein radius of the main lens to be $R_E = 1.0''$, and we place the lens and perturber at redshift $z_{\rm{lens}} = z_p = 0.3$ in front of a source at redshift $z_{\rm{src}} = 2.0$.  We consider three decades of perturber mass.  We generate mock lensing observables and then fit them with Lens+Shear models.  Figure \ref{fig:toyr3} shows the results in terms of scatter in the recovered values of the Hubble constant $h$ and the ellipticity $e$.

The parameter scatter increases as the perturber gets closer to the main galaxy and higher-order terms become more important.  The scalings follow power laws that are consistent with the deviations in the lens potential. Lens+Shear models omit third-order terms and higher, so we expect the scatter to scale as $r^{-3}$. The scatter in the ellipticity and the Hubble constant are both proportional to the mass of the perturbing galaxy. 

Therefore, the scatter in the recovered Hubble constant created by an individual perturbing galaxy is proportional to $M / r^3$ (note that $ M \propto R_p^2$ for a point mass), which is the same scaling as the deviations in the lens potential. These results quantify our intuition that massive galaxies near the main lens are important for lensing. 

Below, we will use these mass and offset scalings to characterize the strength of a perturbing ENV/LOS galaxy, but this current analysis only applies if the perturbing galaxy is at the same redshift as the main lens.

\subsubsection{Foreground Perturbers Are More Important Than Background}
\label{sec:frontback}

Now that we understand the mass and offset scalings, we examine how the redshift of the perturbing galaxy affects its contribution to the lens potential. 

For a single perturbing galaxy, the lens equation \ref{eqn:full_mp} has the form
\begin{equation}
\x_2 = \x_1 - \beta_{1\,2} \al_1(\x_1)
\end{equation}
and
\begin{equation}
\x_s = \x_1 - \beta_{1\,s} \al_1(\x_1) - \beta_{2\,s} \al_2(\x_2).
\end{equation}
Combining these equations, recalling that $\beta_{i\,s} \equiv 1$, and dropping the ``12"" (for convenience) by setting $\beta \equiv \beta_{1\,2}$ lets us write the two-plane lens equation as
\begin{equation}
\x_s = \x_1 - \al_1(\x_1) - \al_2\left(\x_1 - \beta \al_1(\x_1)\right).
\label{eqn:twoplane}
\end{equation}
In the following discussion it is useful to recall that $\beta$ varies monotonically with redshift, ranging from $\beta = 0$ when $z_p = z_{\rm{lens}}$ to $\beta = 1$ when the perturber redshift reaches either $z_p = 0$ (for a foreground perturber) or $z_p = z_{\rm{src}}$ (for a background perturber).

To make further progress, we need to distinguish situations when the perturber is in front of or behind the main lens galaxy. 

\begin{figure}[t]
\begin{center}
\includegraphics[width=\columnwidth]{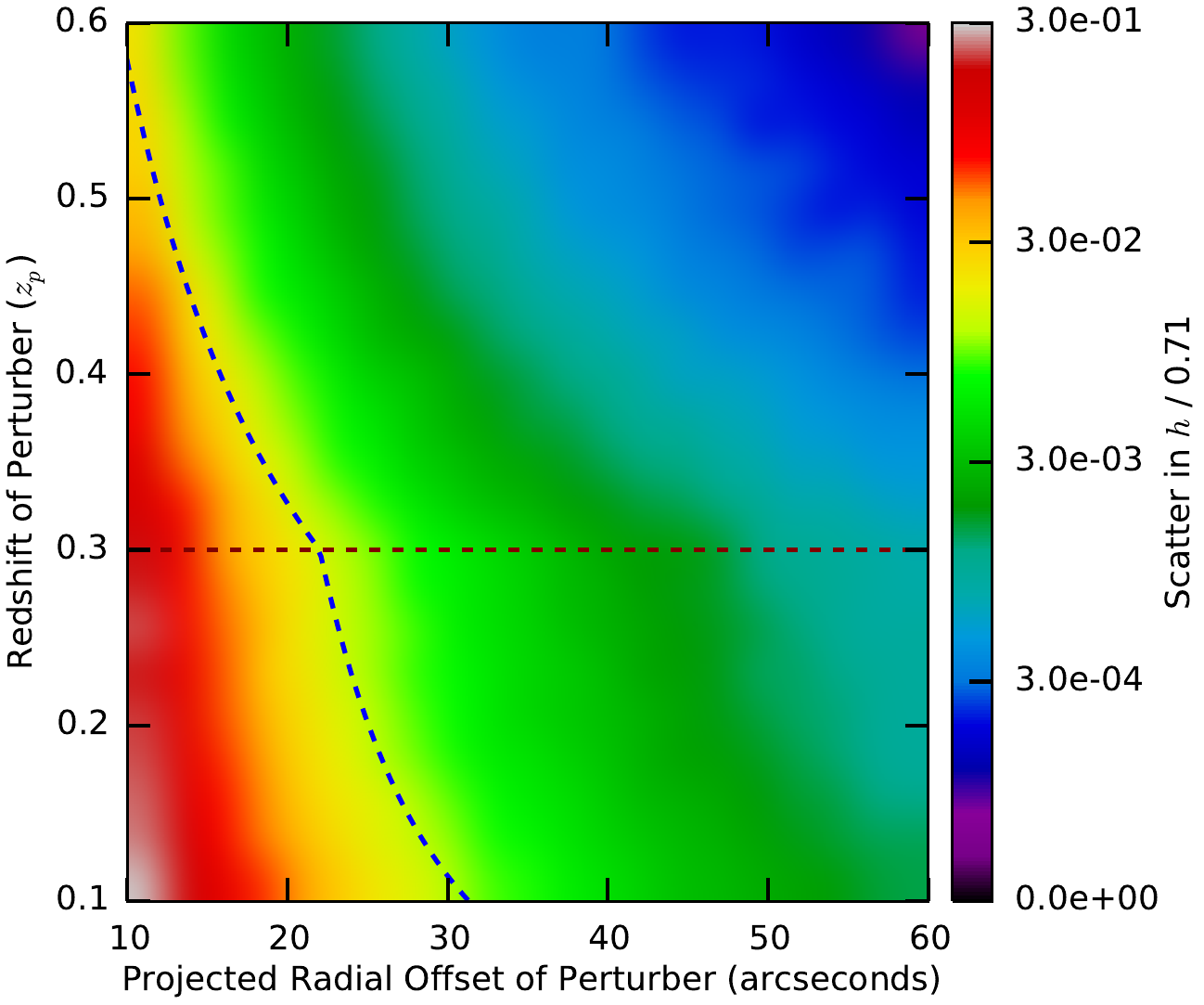}
\caption{\label{fig:toyhd3x} Scatter in the Hubble constant, $h$, recovered from the Lens+3D Tides models, for perturbers at different projected offsets and redshifts. The main lens has a redshift of $z_{\rm{lens}} = 0.3$ which is marked with a dotted line, and the perturbing galaxy has a mass of $10^{12}\,M_\odot$. The dashed line shows a contour of constant flexion shift, $\Delta_3 x$, from equations \ref{eqn:backgroundd3x} and \ref{eqn:foregroundd3x} (picking a different value of $\Delta_3 x$ would simply rescale the contour). The effects due to a perturber in the background are downweighted (see equation \ref{eqn:backgroundd3x}), and therefore the perturber needs to be closer in projection to have the same effect than as if it were at the same redshift as the lens. In the foreground, the curve flares out, implying that foreground perturbers can be farther away in projection to have the same effect if they were at the lens redshift. This occurs because the Einstein radius of a point mass increases as redshift decreases, yielding stronger lensing effects. The curve matches the shape of the color contours, indicating that our theoretical expression from the lens potential captures the redshift dependence of LOS effects in the recovered parameters, like the Hubble constant.  Thus, we can use the flexion shift to compare effects of LOS galaxies even if they are at different redshifts.
}
\end{center}
\end{figure}

\medskip
\centerline{\emph{Lens Potential Perturbations}}
\centerline{\emph{from a Background Galaxy}}
\medskip

Suppose the perturbing galaxy is in the background. The first part of this discussion parallels \citet{Keeton03}.  We have $\al_1 = \al_g$ and $\al_2 = \al_p$. Substituting this into equation \ref{eqn:twoplane} yields
\begin{equation}
\x_s = \x_1 - \al_g(\x_1) - \al_p(\x_1 - \beta \al_g(\x_1)). 
\end{equation}
We Taylor expand $\al_p$ and omit the unobservable terms corresponding to the constant potential and deflection:
\begin{eqnarray}
x^i_s &=& x^i_1 - \alpha^i_g(\x_1) - \Gamma^{ij} (x^j_1 - \beta \alpha^j_g(\x_1)) \\
&&- \frac{1}{2} \sF^{ijk} (x^j_1 - \beta \alpha^j_g(\x_1)) (x^k_1 - \beta \alpha^k_g(\x_1)) + \mathcal{O}(\x_1^4) . \nonumber
\end{eqnarray}
Here we use index notation for clarity.

If we truncate the expansion at tidal terms and return to vector notation, we can write
\begin{equation}
\label{eqn:background}
\x_s = (\I - \GammaMat) \x_1  - (\I - \beta \GammaMat)\al_g(\x_1) + \mathcal{O}(\x_1^3).
\end{equation}
This equation looks very similar to the single plane lens equation, but with some multiplicative factors. In fact, if we multiply both sides by $(\I - \beta \GammaMat)^{-1}$ from the left, introduce a scaled source coordinate $\vec{u}_{\rm eff} \equiv (\I - \beta \GammaMat)^{-1} \x_s$, and define an effective shear by
\begin{equation}\label{eqn:Geff}
\I -\GammaMat_{\rm eff} \equiv (\I - \beta\GammaMat)^{-1} (\I - \GammaMat) ,
\end{equation}
then we can rewrite equation \ref{eqn:background} as
\begin{equation}
\vec{u}_{\rm eff} = (\I - \GammaMat_{\rm eff}) \x_1 - \al_g(\x_1),
\end{equation}
which is equivalent to the single plane lens equation with an external shear in the lens plane \citep[see also][]{Schneider97}. We are allowed to rescale the source plane because the position of the source is unobservable and we typically only measure the ratio of the fluxes of different images, which is insensitive to the absolute intrinsic luminosity of the source.\footnote{The freedom to rescale the source plane vanishes if the intrinsic luminosity of the source is known.  Thus, lensed standard candles, such as type Ia supernovae \citep{Kolatt98, Patel14, Kelly15, Kelly16, Goldstein17, Goobar16,Rodney16}, offer a novel way to break the mass sheet degeneracy.}

In the tidal approximation, we again expect errors in the recovered lens parameters to be associated with the largest terms in the lens potential that have been neglected: the third-order flexion terms.  We therefore use the terms involving $\sF$ to quantify the lens potential perturbations.\footnote{Recall from Section \ref{sec:toylos} that the flexion terms scale with mass and projected offset as $M / r^3$; now we generalize now to include redshift.}  We approximate the main lens galaxy as a singular isothermal sphere (SIS) with deflection
\begin{equation}
\al_g = R_E \rhat.
\end{equation}
Then we can write the lens equation \ref{eqn:background} in the tidal approximation as
\begin{equation}
\label{eqn:backgroundorder2}
\x_s = \x_1 - R_E \rhat -  \GammaMat (\x_1 - \beta R_E \rhat).
\end{equation}
For comparison, the lens equation with flexion terms is
\begin{multline}
\x_s = \x'_1 - R_E\rhat -  \GammaMat(\x'_1 - \beta R_E\rhat) \\
- \frac{1}{2}(\x'_1 - \beta R_E\rhat) \sF (\x'_1 - \beta R_E \rhat).
\label{eqn:backgroundorder3}
\end{multline}
where we recognize that the solutions $\x'_1$ of this equation may differ from the solutions $\x_1$ of equation \ref{eqn:backgroundorder2} because of the flexion effects.  If we assume that $\GammaMat \ll 1$, and we subtract equations \ref{eqn:backgroundorder3} and \ref{eqn:backgroundorder2}, we can define the image shift caused by third-order terms to be $\Delta_3 \x \equiv \x'_1 - \x_1$, and thus find
\begin{equation}
\Delta_3 \x = \frac{1}{2} (\x'_1 - \beta R_E \rhat) \sF (\x'_1 - \beta R_E \rhat).
\end{equation}
Now, if we define the perturber to be collinear with the image position ($\theta  = 0$)\footnote{If we consider all possible angles, the root mean square value has an extra factor of $\sqrt{2}$ that does not affect the scalings.} and assume that the positions of the multiple images are $|\x'_1| \approx |\x_1| \approx R_E$, we can write the magnitude of the flexion shift as
\begin{equation}
\Delta_3 x = \frac{R_E^2 R_p^2}{r^3} (1 - \beta)^2. 
\label{eqn:backgroundd3x}
\end{equation}
We call this quantity, $\Delta_3 x$, the ``flexion shift'' because it characterizes the perturbations to the image positions caused by the third-order (``flexion'') terms from a perturber. This quantity has units of arcseconds and it gives us a way to quantify and compare the ENV/LOS contributions from different perturbing galaxies.

We note that in this derivation, we have used the expression for $\sF$ for a point mass, which is reasonable when the perturber is projected far from the images.  Strictly speaking, $r$ here is the unlensed distance in the redshift plane of the perturber. To convert to offset as observed on the sky $r'$, we use the lens equation to yield $r' = r - \beta R_E$. As long as the perturbing galaxy is many Einstein radii away from the lens, we can put $r' \approx r$.  (If that were not the case, the perturber should probably not be treated using the tidal approximation anyway.)

\medskip
\centerline{\emph{Lens Potential Perturbations}}
\centerline{\emph{from a Foreground Galaxy}}
\medskip

We now turn our attention to a perturber in the foreground of the lens. In this case $\al_1 = \al_p$ and $\al_2 = \al_g$, so the lens equation has the form
\begin{equation}
\x_s =\x_1 - \al_p(\x_1) - \al_g(\x_1 - \beta \al_p(\x_1)).
\end{equation}
Taylor expanding $\al_p$ yields
\begin{eqnarray}
x^i_s &=& x^i_1  - \alpha_p^i(0) - \Gamma^{ij}x^j_1 - \frac{1}{2} \sF^{ijk} x^j_1 x^k_1 +\ldots \\
&& - \alpha^i_g \left(x^i_1 - \beta\alpha_p^i(0) - \beta \Gamma^{ij} x^j_1 - \frac{1}{2} \sF^{ijk} x^j_1 x^k_1 + \ldots \right). \nonumber
\end{eqnarray}
Now truncating at tidal terms yields
\begin{equation}
\x_s = (\I - \GammaMat) \x_1 - \al_g\left((\I - \beta\GammaMat) \x_1 + \mathcal{O}(\x_1^3)\right) + \mathcal{O}(\x_1^3).
\label{eqn:foregroundlenseqn}
\end{equation}
 
We now seek an expression for the flexion shift $\Delta_3 x$ caused by a foreground perturber to compare to the background case.  Following an analysis similar to that above, and again using an SIS main lens and a point mass perturber, we can write the lens equation in the tidal approximation as
\begin{equation}
\x_s = \x_1 - R_E \rhat - \GammaMat(\x_1 - \beta R_E \rhat) 
\end{equation}
while the lens equation with flexion terms is
\begin{equation}
\x_s = \x'_1 - R_E \rhat - \GammaMat(\x_1 - \beta R_E \rhat) - \frac{1}{2} \x_1 \sF \x_1.
\end{equation}
Subtracting these equations and rearranging yields
\begin{equation}
\Delta_3 \x = \frac{1}{2} \x_1 \sF \x_1.
\end{equation}
In this case, the multiple images form at $\x_2 \approx R_E \rhat$, implying that $R_E \approx \x_1 - \beta \GammaMat \x_1$. If we assume that $\GammaMat \ll 1$ (as above), we can write $\x_1 \approx R_E \rhat$, giving
\begin{equation}
\Delta_3 x = \frac{R_E^2 R_p^2}{r^3}.
\label{eqn:foregroundd3x}
\end{equation}
This equation is similar to equation \ref{eqn:backgroundd3x}, but without any $\beta$ factors. Therefore, the effects of background perturbers are downweighted compared to foreground perturbers.

\medskip
\centerline{\emph{Scatter in Lens Model Parameters}}
\centerline{\emph{Tracks Lens Potential Perturbations}}
\medskip

We now use simulations to test whether our expression for the deviation in the lens potential, the flexion shift, $\Delta_3 x$ (eqs.\ \ref{eqn:backgroundd3x} and \ref{eqn:foregroundd3x}), provides a useful way to characterize the importance of ENV/LOS perturbers and their contribution to the recovered lens parameters and Hubble constant. For these simulations, we again adopt the simplified mass distribution of a main lens galaxy treated as an SIS with Einstein radius of 1'' at $z=0.3$ and a single point mass. We fix the perturber mass to be $10^{12}\,M_\odot$ but vary its projected offset and redshift. In these simulations, the assumption of treating the perturber as a point mass is not valid when the projected offset or redshift is small, but is reasonable in the regime where we anticipate using the tidal approximation.  We use the exact lens equation to generate mock lensing observables, and then fit them using Lens+3D Tides models.  Figure \ref{fig:toyhd3x} shows the scatter in the recovered value of the Hubble constant (the median value is not shown because it matches the input value; there is no bias).  Generally speaking, the scatter increases as the offset decreases and higher order terms become more significant. In the background, the perturber's effects become weaker as the redshift offset increases because of the $\beta$ factors in equation \ref{eqn:backgroundd3x}; while in the foreground, the perturber's effects become stronger as $z \to 0$ because its angular Einstein radius increases.  Both of these scalings are well described by our functional form for $\Delta_3 x$, which indicates that we have successfully defined a quantity to characterize the contribution of perturbing galaxies. 

Many lens systems have significant contributions from in their local environment; for example, HE0435$-$1223 has a neighbor \citep{Kochanek06} while MG0414+0534 \citep{Tonry99}, RXJ1131$-$1231 \citep[hereafter RXJ1131,][]{Sluse03}, and B2114+022 \citep{King99} all have satellites that are presumably close enough to matter. While some models do treat nearby perturbers exactly \citep[e.g.,][]{Kochanek06, Fadely12}, typically the decision about whether to include a neighbor galaxy in a model is \textit{ad hoc}. The flexion shift, $\Delta_3 x$, provides a quantitative criterion to rank order the importance of LOS galaxies even if they are at different redshifts. Any given perturber can be treated with the tidal approximation if its $\Delta_3 x$ value is small, but it must be treated explicitly if its $\Delta_3 x$ value is large (see \citealt{Sluse16} and \citealt{Rusu16} for applications of the flexion shift to observed LOS galaxy catalogs).

\medskip
\centerline{\emph{Foreground Galaxies Produce}}
\centerline{\emph{nonlinear Lens Potential Perturbations}}
\medskip
While the lens equation for a foreground perturbing galaxy (eq.\ \ref{eqn:foregroundlenseqn}) looks similar to that of a background perturber (eq.\ \ref{eqn:background}), there is a key difference: instead of just having a multiplicative effect on the source position like the background perturber, the deflection from the foreground perturber enters the lens equation \emph{inside the argument} of the deflection of the main galaxy. A foreground perturber produces nonlinear effects because it creates a difference between the coordinates we see on the sky and the coordinates in the plane of the main lens galaxy, where we need to evalute the lensing potential. There is no way to define an effective shear that fully captures the nonlinear effects of a foreground perturber (see also M14).

In principle, one can define a scaled coordinate based on the argument of the deflection to transform this equation to look like the standard lens equation \citep[e.g.,][]{Schneider97,Keeton03}. This requires care, however, because the new quantities do not correspond to the \emph{observed} image positions that are typically used in lens modeling. An alternative is to rescale the mass of the main lens galaxy to account for the change in the argument of the deflection \citep{Schneider97}, but then the mass predicted by the lens models longer corresponds to the true physical mass of the lens. 

\begin{figure}[t]
\begin{center}
\includegraphics[width=1.0\columnwidth]{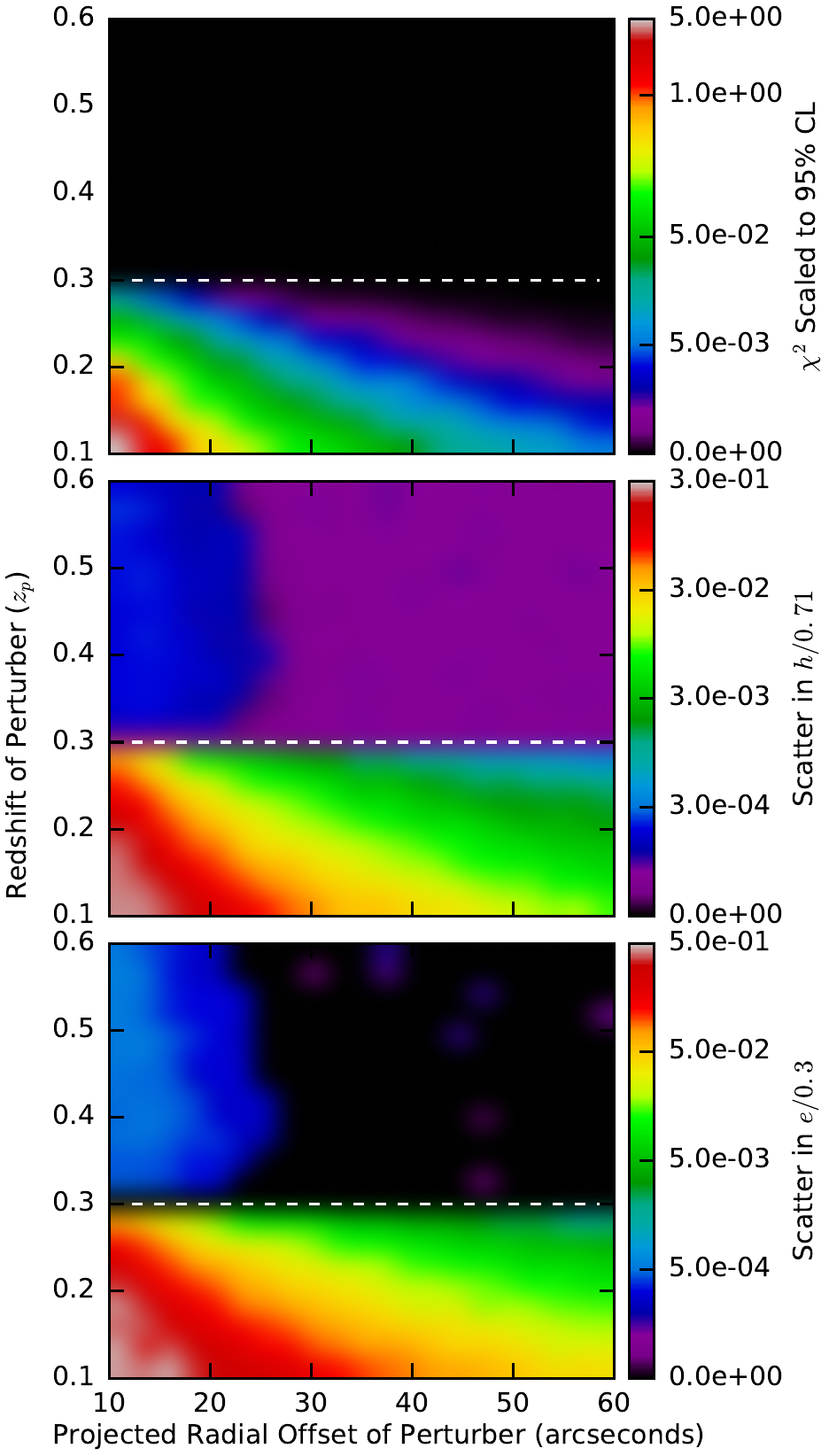}
\caption{\label{fig:frontback} Recovered lens model parameters for the Lens+Shear models from simulations with only tidal effects from an LOS perturber. The simulations shown here do not include higher-order terms, to emphasize nonlinear effects.  The main lens has a redshift of $z_{\rm{lens}} = 0.3$ (marked by the white dashed line) and the perturber has a mass of $10^{12}\, M_\odot$.  We use an $\asinh$ color scaling for dynamic range; (at large values, $\asinh$ acts like a logarithm, but at small values, it becomes linear).  When the perturber is in the background, an effective shear can account for the LOS effects (see eq.\ \ref{eqn:Geff}).  When the perturber is in the foreground, however, there are nonlinear effects that cannot be mimicked by a shear in the lens plane.
}
\end{center}
\end{figure}

To examine the differences between the foreground and background pertubers, we again use a simplified mass distribution of a single $10^{12} M_\odot$ perturbing galaxy at a variety of projected offsets and redshifts and a 1'' SIS main lens galaxy. To emphasize the contribution from nonlinear effects, we also generate the mock lensing observables in a simplified way, using the tidal approximation (i.e., there are no higher-order terms), but otherwise we follow the procedure from Section \ref{sec:methods}. This allows us to focus entirely on nonlinear effects.  Figure \ref{fig:frontback} shows the scatter in the recovered parameters from the Lens+Shear models. An external shear can account for a background perturber (to within numerical precision), as we expect from our analysis of the lens equation for a background perturbing galaxy (eq.\ \ref{eqn:background}).  By contrast, shear \emph{cannot} mimic the nonlinear effects of a foreground LOS galaxy. 

\subsection{Characterizing Realistic Beams\label{sec:Environments}}

\begin{figure*}[t]
\begin{center}
\includegraphics[width=1\textwidth]{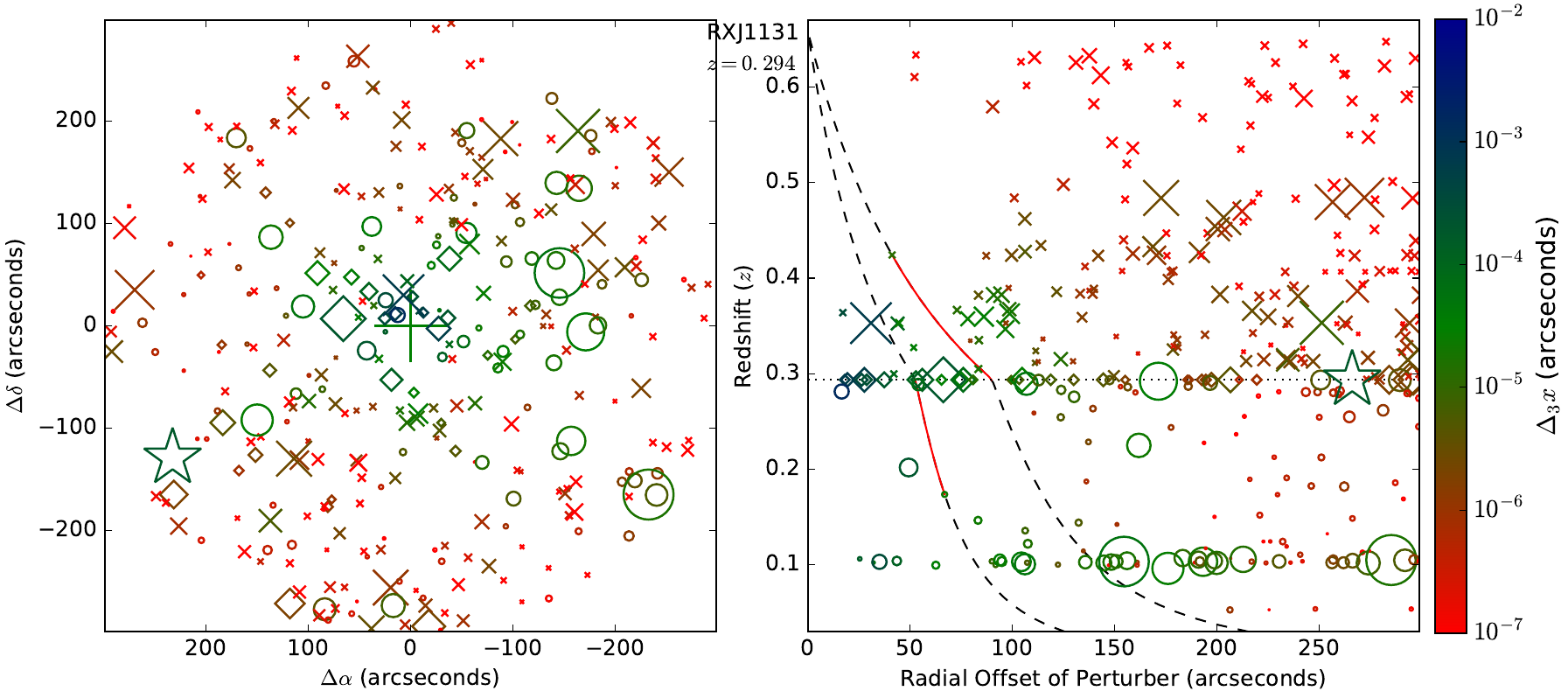}
\caption{\label{fig:fieldrz} Illustration of the flexion shift for galaxies in the field of RXJ1131.  In the left panel, each galaxy is shown at its position on the sky.  The area of each point is proportional to the mass of the galaxy.  X and O symbols mark galaxies behind and in front of the main lens galaxy, respectively, while diamonds mark members in the group around the main lens, and the star indicates the location of the common group halo.  The color of the points represents the strength of the lensing effects measured by the flexion shift $\Delta_3 x$. The right panel shows the same galaxies plotted in the $r$--$z$ projection, similar to Figures \ref{fig:frontback} and \ref{fig:toyhd3x}. The main lens redshift is indicated by the dotted line. Two $\Delta_3 x$ contours have been shown to guide the eye. The red section of the contour illustrates how to map a perturbing galaxy to its effective offset as if it were in the main lens plane. There is no simple radial cut that selects the most important LOS galaxies; a more complicated quantity like the flexion shift, $\Delta_3 x$, is necessary.%
}
\end{center}
\end{figure*}

\begin{figure}
\begin{center}
\includegraphics[width=1\columnwidth]{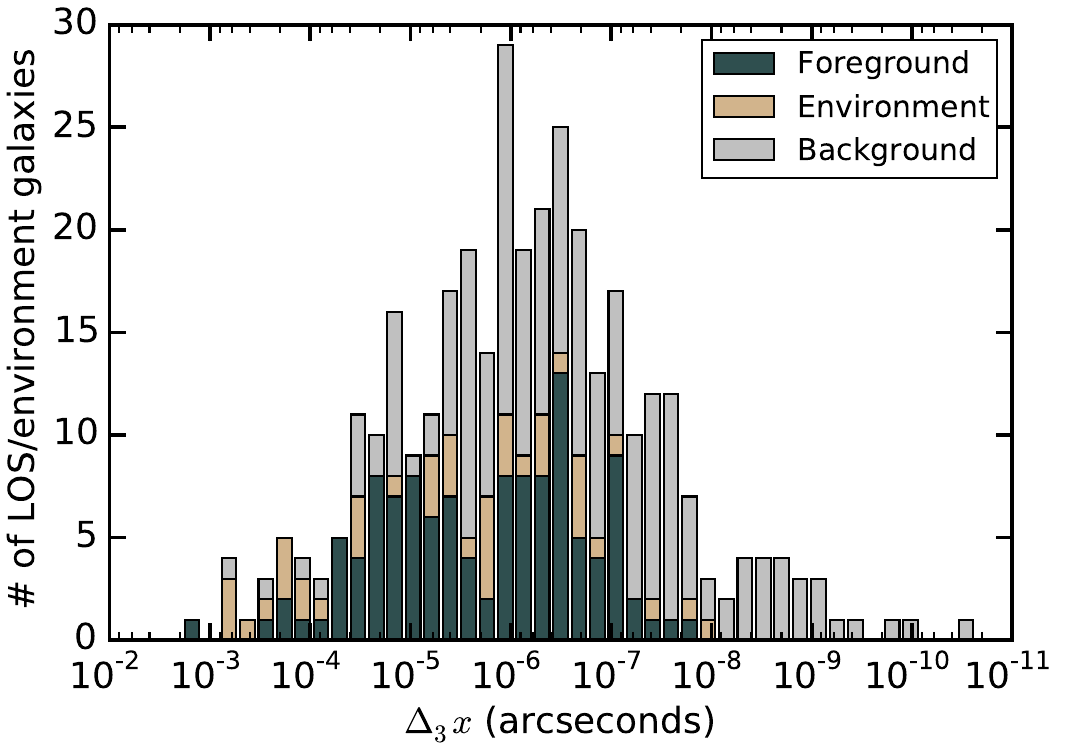}
\caption{\label{fig:d3xhist} Histogram of flexion shifts for galaxies within 5' of RXJ1131.  The dark (light) gray bins represent perturbing galaxies in the foreground (background) of the main lens galaxy, while the tan bins in between correspond to group members in the immediate environment of the main lens. The range of flexion shifts for this field spans $\sim 8$ orders of magnitude. We have reversed the x-axis for comparison to subsequent figures. Perturbers with large a flexion shift, and therefore a large contribution to the lens potential, are on the left. For reference, when generating our $\chi^2$ values, we have assumed a positional uncertainty of $3 \times 10^{-3}$ arcsec. Few individual perturbing galaxies have a flexion shift that large, but the cumulative flexion is easily larger than our assumed measurement uncertainties. As anticipated, foreground perturbers and group members tend to have larger $\Delta_3 x$ values than background perturbers.}
\end{center}
\end{figure}

\begin{figure}[t]
\begin{center}
\includegraphics[width=\columnwidth]{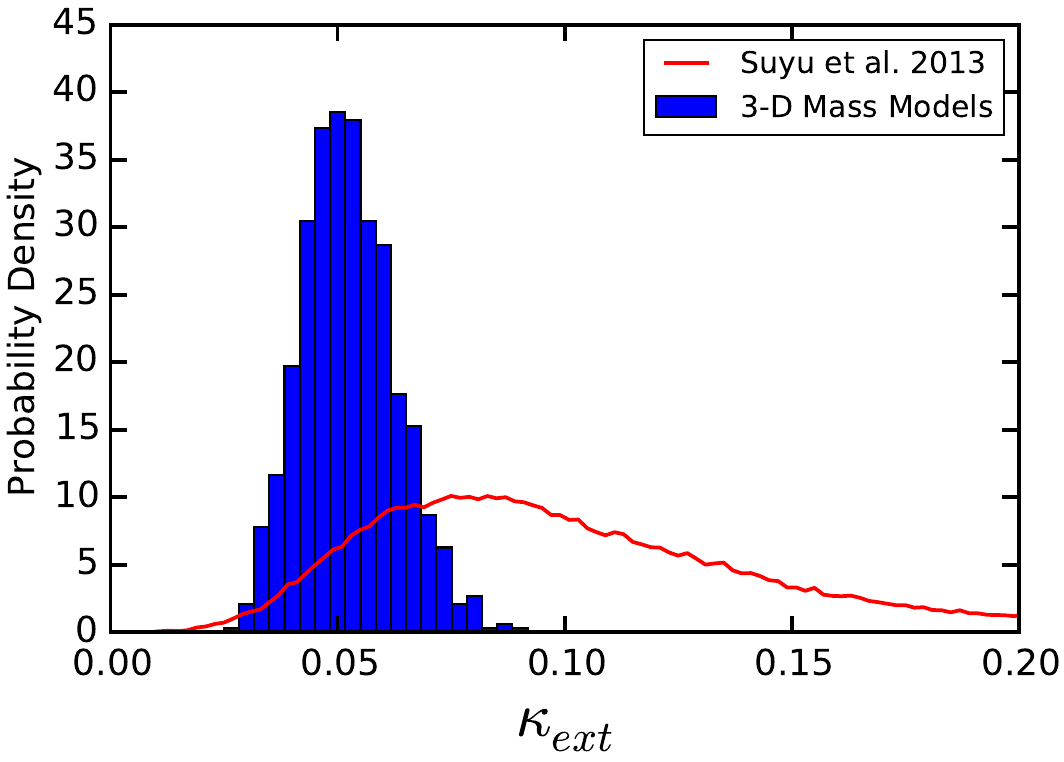}
\caption{\label{fig:kappa} Probability distribution for the external convergence in the RXJ1131 field. The blue histogram shows the effective convergence computed directly from our 3D mass models, accounting for measurement uncertainties and scatter in the relations used to assign masses to galaxies. The red curve shows the external convergence distribution derived using galaxy number counts from the Millennium simulation by \citet{Suyu13}. The peaks of the distributions are roughly consistent, but our models favor slightly lower values of external convergence. Because we build mass models for the specific, observed field, we obtain a $\sim 4$ times narrower distribution that effectively translates into a stronger prior on the Hubble constant.%
}
\end{center}
\end{figure}

\begin{figure*}
\begin{center}
\includegraphics[width=1\textwidth]{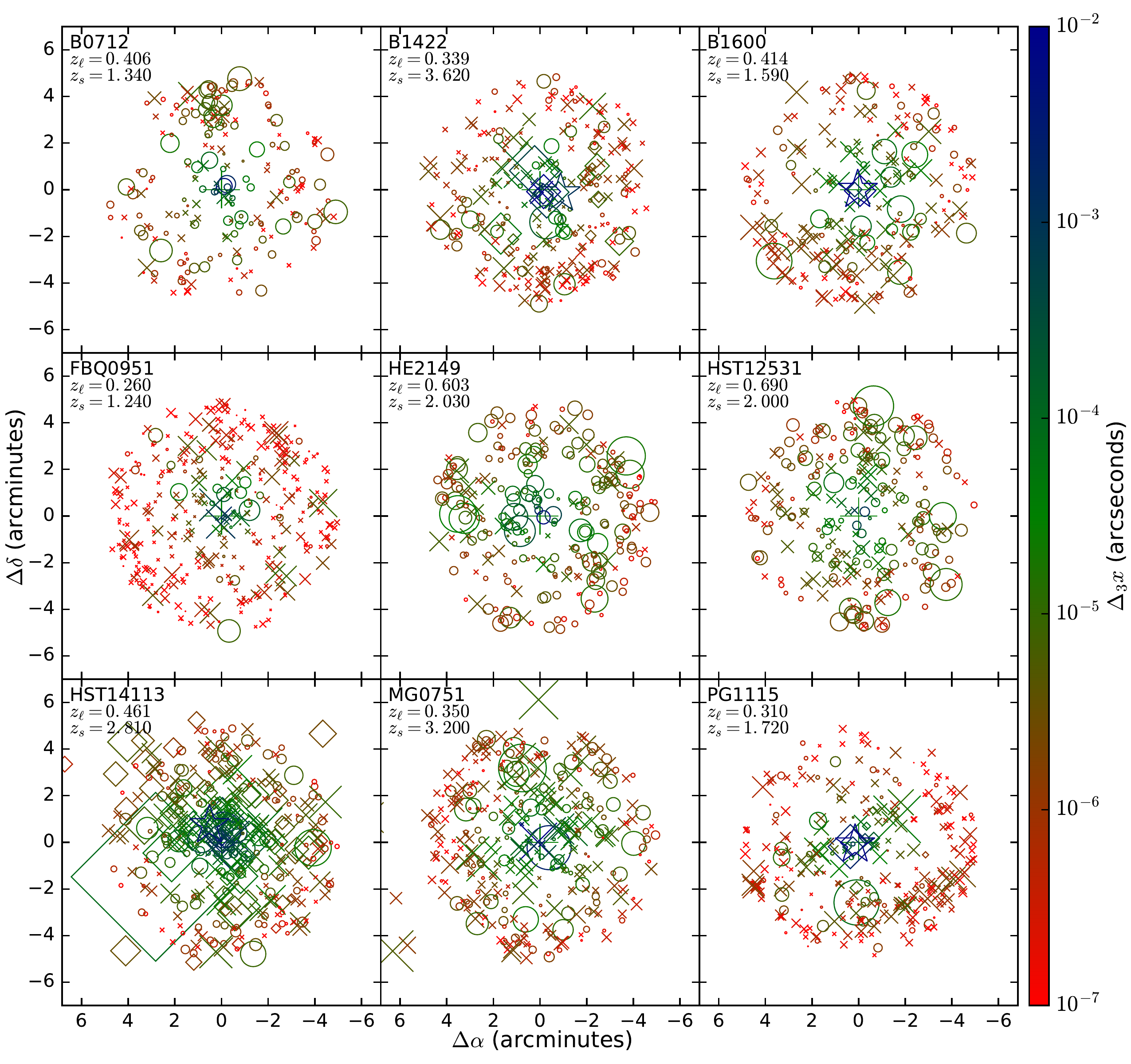}
\caption{\label{fig:allfields} Illustration of the flexion shifts, $\Delta_3 x$, for perturbing galaxies in nine lens fields (similar to the left panel in Fig.\ \ref{fig:fieldrz}). Each galaxy in the mass model is shown in projection on the sky. The area of each point is proportional to the mass of the galaxy. X symbols are behind the main lens galaxy, while O symbols are in the foreground. Diamonds represent group members, while stars indicate the locations of group halos. The color of the points represents the strength of the higher-order terms characterized by the flexion shift, $\Delta_3 x$. There is dramatic variation in the ENV/LOS strengths across the different fields, implying that each field needs to be treated on an individual basis. For all of the fields shown here, there is no simple radial cut that would capture all of the important galaxies: a more complicated quantity like the flexion shift is necessary. 
}
\end{center}
\end{figure*}

Now that we have our quantity to assess the importance of individual galaxies, we explore what we can learn from the realistic matter distributions alone (that are built following Section \ref{sec:massmodels}), even before we fit mock lensing data. Our primary use case for the flexion shift is to rank order which of the $sim$ few hundred ENV/LOS galaxies produce the strongest effects on the inferred Hubble constant, which we will treat more carefully (i.e., as ``exact'' planes in the 3D Lens models). The flexion shift criterion also gives us the ability to compare the mass distributions for different lens beams. This gives us a quantitaive way to select fields that do not have strong ENV/LOS contributions as a ``gold sample'' of lenses for future surveys like LSST and Euclid. 

To gain intuition about how the flexion shift applies to realistic beams, we first consider the well studied lens system RXJ1131. We adopt the measured source and lens redshifts of $z_{\rm{src}} = 0.658$ and $z_{\rm{lens}} = 0.2936$, and a lens ellipticity of $e=0.3$, and an Einstein radius of $R_E = 1.5\arcsec$, which is close to the measured value \citep{Suyu13} and also close to the peak of the observed distribution \citep{Sonnenfeld13}. Following Section \ref{sec:massmodels}, we produce the matter distributions from our observations of RXJ1131 and then calculate the flexion shift for all of the perturbing ENV/LOS galaxies. Figure \ref{fig:fieldrz} illustrates $\Delta_3 x$ values for galaxies in this field.  While $\Delta_3 x$ generally increases toward the lens, there is no single radial cut that can be used to determine the importance of a perturber because perturber mass and redshift must also be taken into account.  By following contours of constant $\Delta_3 x$, as illustrated in the right-hand panel, we can determine the effective radial offset for the perturber if it were in the main lens plane. Since background perturbers are downweighted by ($1-\beta)^2$, they must have a smaller projected offset to have the same effect as if they were in the main lens plane. Perturbing galaxies in the foreground are slightly upweighted because the Einstein radius of a point mass perturber increases as we move it to lower redshift, which means that foreground galaxies can have a larger projected offset to have the same effect as if they were at the main lens redshift.

Figure \ref{fig:d3xhist} shows a histogram of all the $\Delta_3 x$ values for this field.  The flexion shifts range from $\sim 5 \times 10^{-3}$ arcsec down to $<10^{-10}$ arcsec, with background perturbers generally having smaller values than foreground perturbers or group members.  In the next section we consider the quantitative threshold for $\Delta_3 x$ needed to achieve desired accuracy and precision in lens model results.

While external convergence does not capture all LOS effects (it omits nonlinearities in eq.\ \ref{eqn:full_mp}), it is a useful quantity for comparing to previous work. \citet[][see also \citealt{Rusu16}]{Suyu13} used the Millennium simulation along with galaxy number counts around RXJ1131 to estimate a prior probability distribution for the external convergence used to constrain the Hubble constant; their distribution is shown in red in Figure \ref{fig:kappa}. We compute the effective convergence directly from our mass models for RXJ1131, as defined in M14 and references therein as
\begin{equation}
\GammaMat_{\rm{eff}} = \I - \C^{-1}_{\ell s} \B_s \B^{-1}_\ell.
\end{equation}

We consider many realizations that account for observational uncertainties and scatter in the relations used to assign masses to galaxies described in Section \ref{sec:massmodels}.  This yields a distribution of ``direct'' external convergence values that is shown by the blue histogram in Figure \ref{fig:kappa}. We find that the peaks of the distributions are roughly consistent. However, our ``direct'' convergence calculations produce a tighter distribution by a factor of $\sim 4$. This is likely because we build mass models based on the observed beam for each individual lens rather than using statistical results from $N$-body simulations. The narrower distribution of external convergences translates into a stronger prior on measuring the Hubble constant.

We can also characterize values of the flexion shift for a wider range of fields.  Figure \ref{fig:allfields} shows $\Delta_3 x$ values for nine other multiply imaged QSO fields from \citet{Wong11}.  The fields are all complex; there is not a simple radial cut that divides the sample into exact and tidal perturbers.  There is striking diversity in the strength of external effects.  There is no single number of galaxies that is guaranteed to be the ``right'' number to include in lens models.  Each field must be considered on an individual basis.

\begin{figure}[!t]
\centering
\includegraphics[width=\columnwidth]{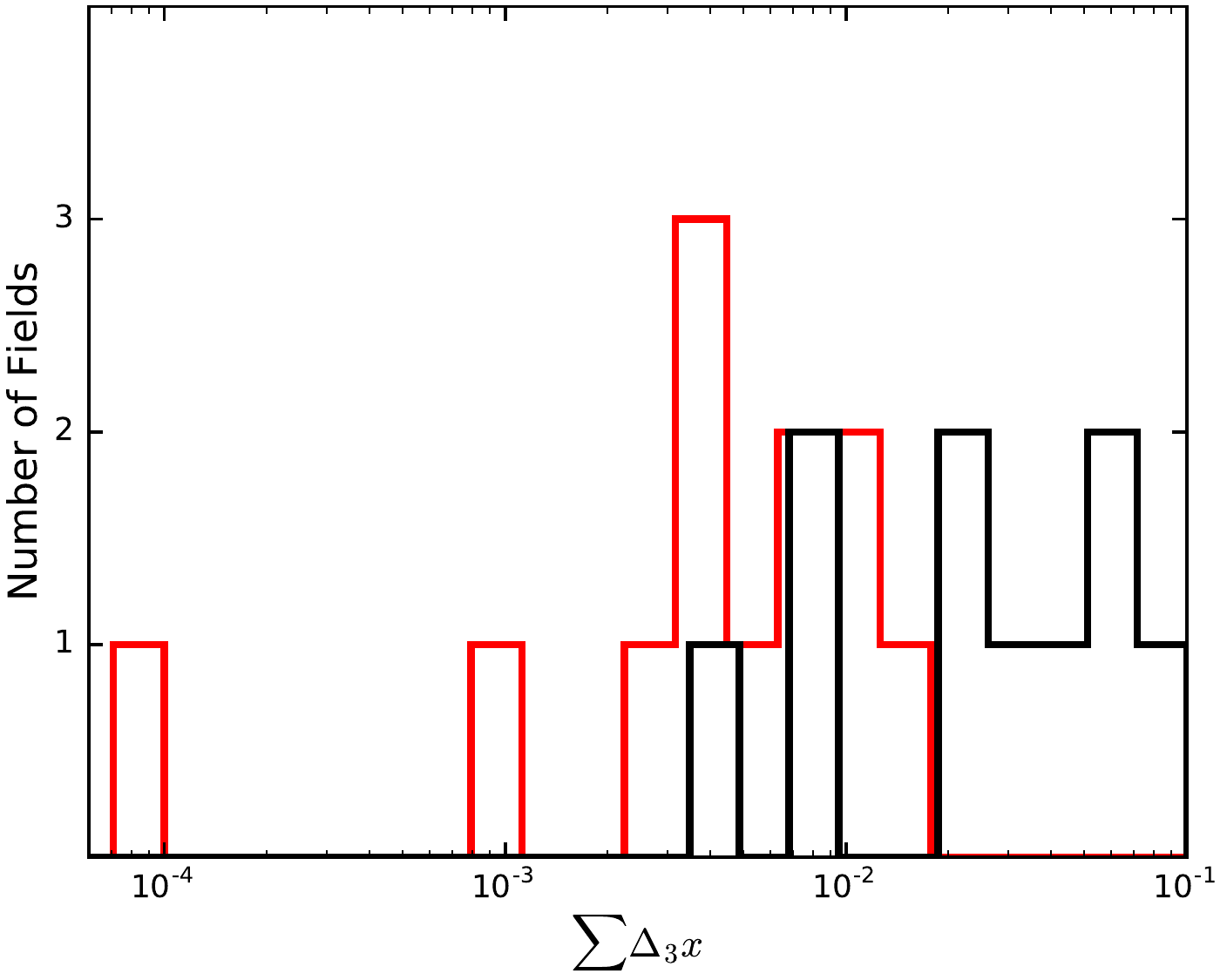}
\caption{Distribution of ENV/LOS strengths for our sample of 3D mass models. Beams for which the lens galaxy is a member of a group are shown in black, while non-group lenses are shown in red. Lenses in groups typically have stronger ENV/LOS contributions than those that are not group members. However, there is substantial overlap so each field needs to be treated individually.}
\label{fig:d3xsums}
\end{figure}

While each field is unique, we do expect that some fields will have stronger ENV/LOS contributions than others. One important factor may be whether the main lens galaxy is part of a group. To make a simple quantitative comparison, we add the flexion shifts ($\Delta_3 x$ values) in a given beam. The higher-order effects due to flexion terms combine in a more complicated way (as the flexion is a tensor), but the simple sum gives an a good estimate of the importance of the ENV/LOS of the lens. If there are many nearby, massive perturbers then the sum will be large, implying a strong ENV/LOS contribution (such as HST14113 in Figure \ref{fig:allfields}). Conversely, in a sparser field the sum will be smaller (such as B0712). Figure \ref{fig:d3xsums} shows a histogram of the sums for our set of 23 lens fields. While the sample size is not large, it is apparent that lenses that are members of a group tend to have stronger ENV/LOS contributions than those that are not.

\subsection{Fitting Lensing Observables}
\label{sec:fitting}
Now that we have characterized the mass distributions for a variety of lens systems, we consider whether different methods for treating ENV/LOS effects can reliably recover lens galaxy parameters and the Hubble constant. We generate mock lensing observables using the full 3D mass models in the simulations described in Section \ref{sec:observables}; throughout this analysis we combine results for the 300 mock quad lens systems with different source positions and azimuthal angles. We then fit the mock lensing data using the types of models discussed in Section \ref{sec:lensmodels}. We first test whether simple Lens-Only and Lens+Shear models can produce robust constraints on the Hubble constant. We then examine the Lens+3D Tides models to ascertain the importance of nonlinear effects. Finally, we add back in some of the higher-order terms for the general 3D Lens models to test when the tidal approximation is applicable and we consider the parameter degeneracies to search for the fundemental limitations in recovering cosmological parameters from lens modeling in general.

\begin{figure*}[ht]
\begin{center}
\includegraphics[width=0.95\textwidth]{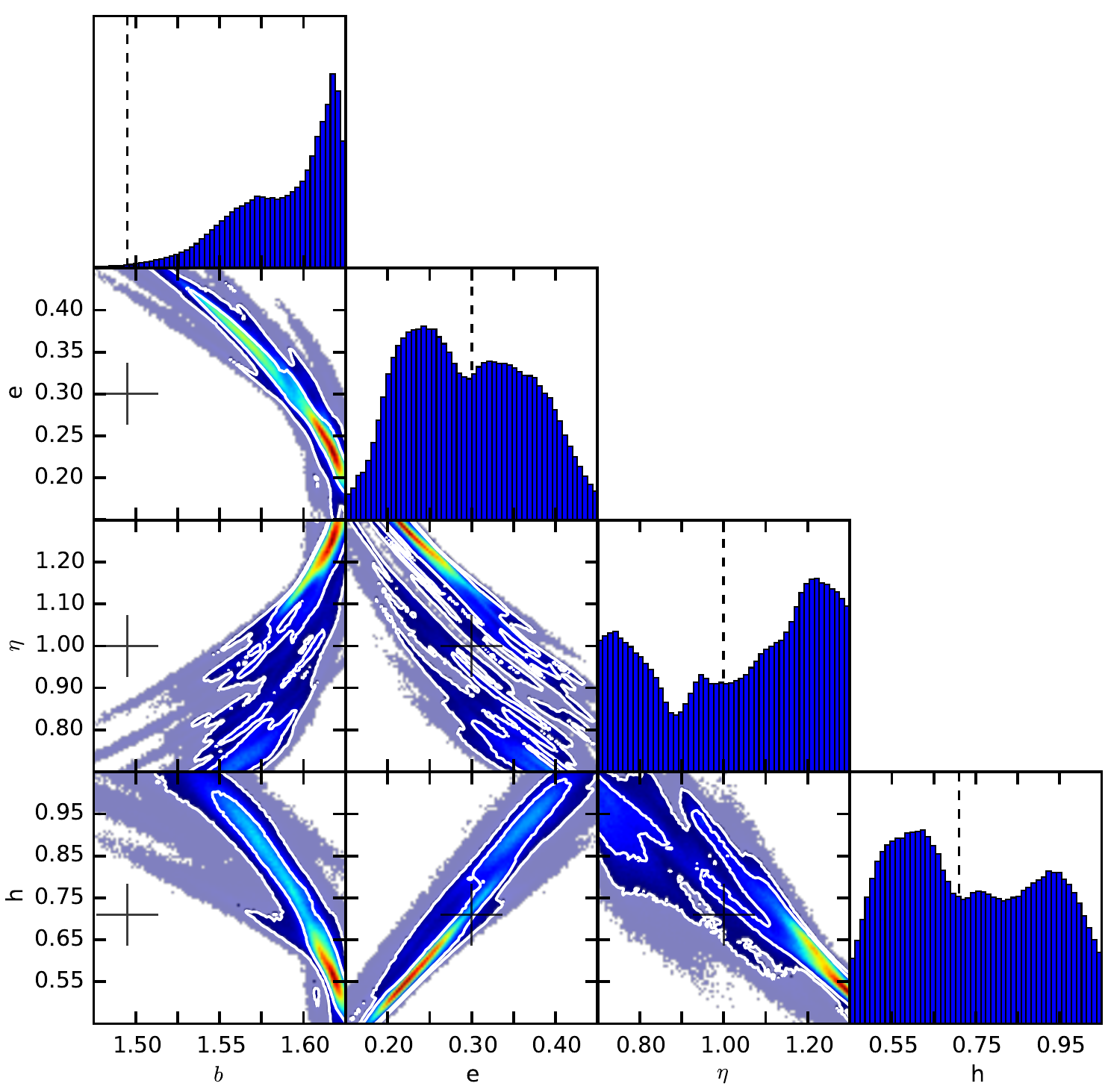}
\caption{\label{fig:none_triangle} Posterior distributions for model parameters from Lens-Only models fit to mock quad lenses in the field of RXJ1131. The true input values are marked with crosses in the contour plots and with a dashed line in the histograms. We combine MCMC results for 300 mock lenses with different source positions and azimuthal angles for the main lens galaxy.  Contours containing $68\%$ and $95\%$ of the MCMC trials are shown in white. Many of the parameters show significant biases and the distributions are often degenerate and multi-modal. These results suggest that ENV/LOS cannot be ignored. The recovered values of the Hubble constant have a scatter of $\sim 20\%$, more than an order of magnitude larger than our goal for ``Precision Lensing'' of $<1\%$. The plotting ranges for corresponding parameters are the same as Figures \ref{fig:shear_triangle} and \ref{fig:los_triangle} to facilitate direct comparison.
}
\end{center}
\end{figure*}

We begin with Lens-Only models that do not explicitly account for any ENV/LOS effects (Section \ref{sec:lensonly}). Figure \ref{fig:none_triangle} shows the recovered lens model parameters including the mass normalization (which is related to the Einstein radius; see eq.\ \ref{eqn:powerlaw}), ellipticity, power law index, and Hubble constant for the RXJ1131 field. The posterior distributions of the recovered lens parameters are broad and often multi-modal. All of the parameters show significant bias, and are not able to reproduce the input parameters. These results reiterate (not surprisingly) that ENV/LOS effects cannot be ignored.

\begin{figure*}[ht]
\begin{center}
\includegraphics[width=1\textwidth]{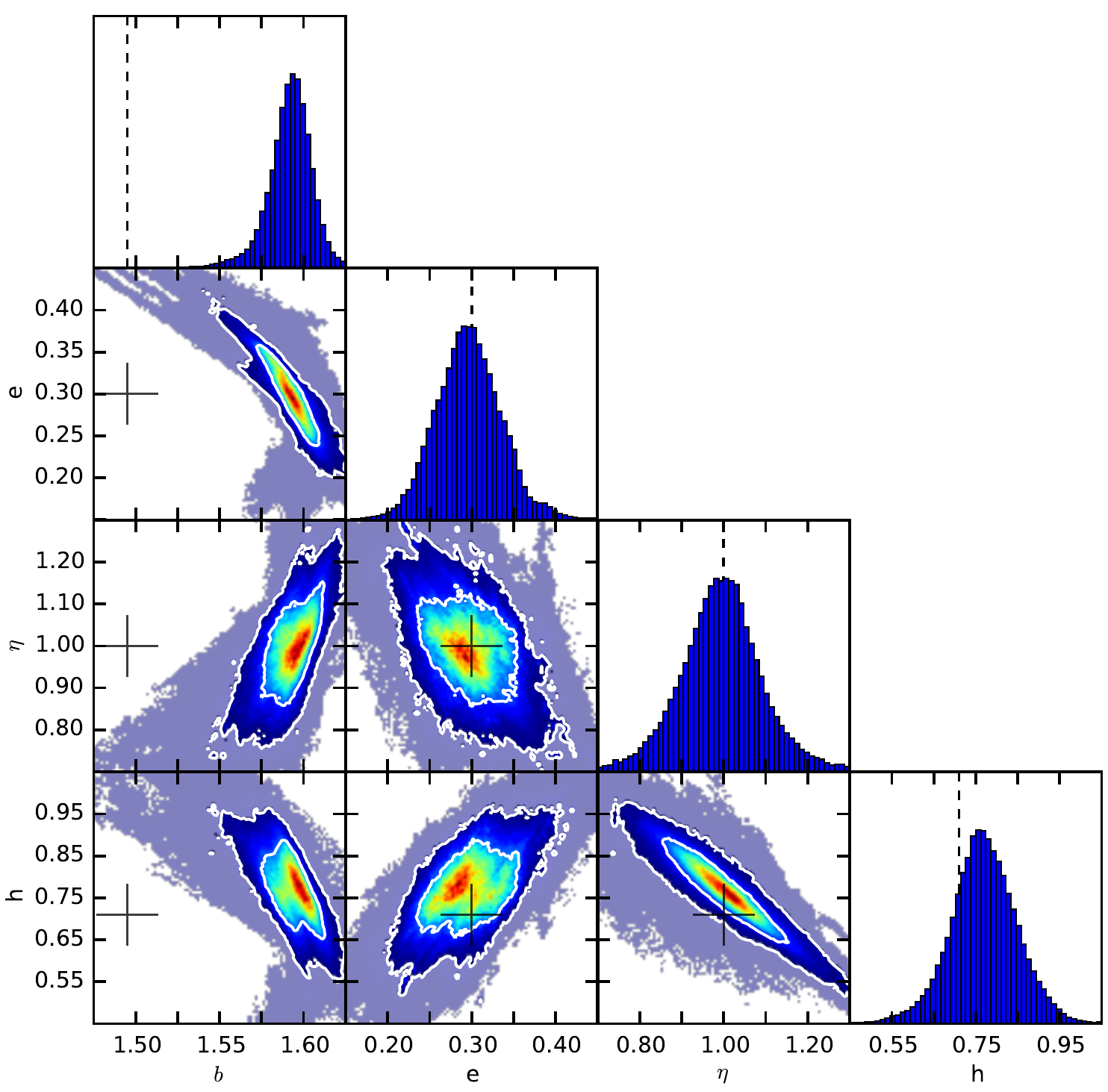}
\caption{\label{fig:shear_triangle} Similar to Figure \ref{fig:none_triangle}, but for Lens+Shear models. Including an external shear dramatically improves the fits over ignoring the ENV/LOS, but the distribution of recovered parameters are non-Gaussian and may have multiple families of solutions. Many of the parameters show biases; most notably, the Einstein radius parameter ($b$) is strongly biased because Lens+Shear models do not explicitly include external convergence (such models must add external convergence in post-processing).  
}
\end{center}
\end{figure*}

Next we consider the Lens+Shear models (Section \ref{sec:lens+shear}. These models have two extra free parameters that can be used to capture some ENV/LOS effects. The results of the fits are shown in Figure \ref{fig:shear_triangle}. The posterior distributions of the recovered lens parameters are considerably tighter than for Lens-Only models. However, these models still have biases that are most dramatic in the Einstein radius and Hubble constant. The bias is associated with external convergence: the total mass in the simulated lenses includes contributions from the environment and LOS; if we assign all of that mass to the main lens galaxy (as in Lens+Shear and Lens-Only models), we overestimate the galaxy's mass and Einstein radius.  In this approach, external convergence must be included through post-processing \citep{Collett13, Suyu10}. The biases that we find are larger than those found in \citet{Jaroszynski16} for random sight lines in the Millenium Simulation. However, they point out that their sight lines, on average, are not as dense as that of RXJ1131. This is in agreement with \citet{Collett16} who find that lenses tend to be in (slightly) overdense lines-of-sight.

\begin{figure}[ht]
\centering
\includegraphics[width=\columnwidth]{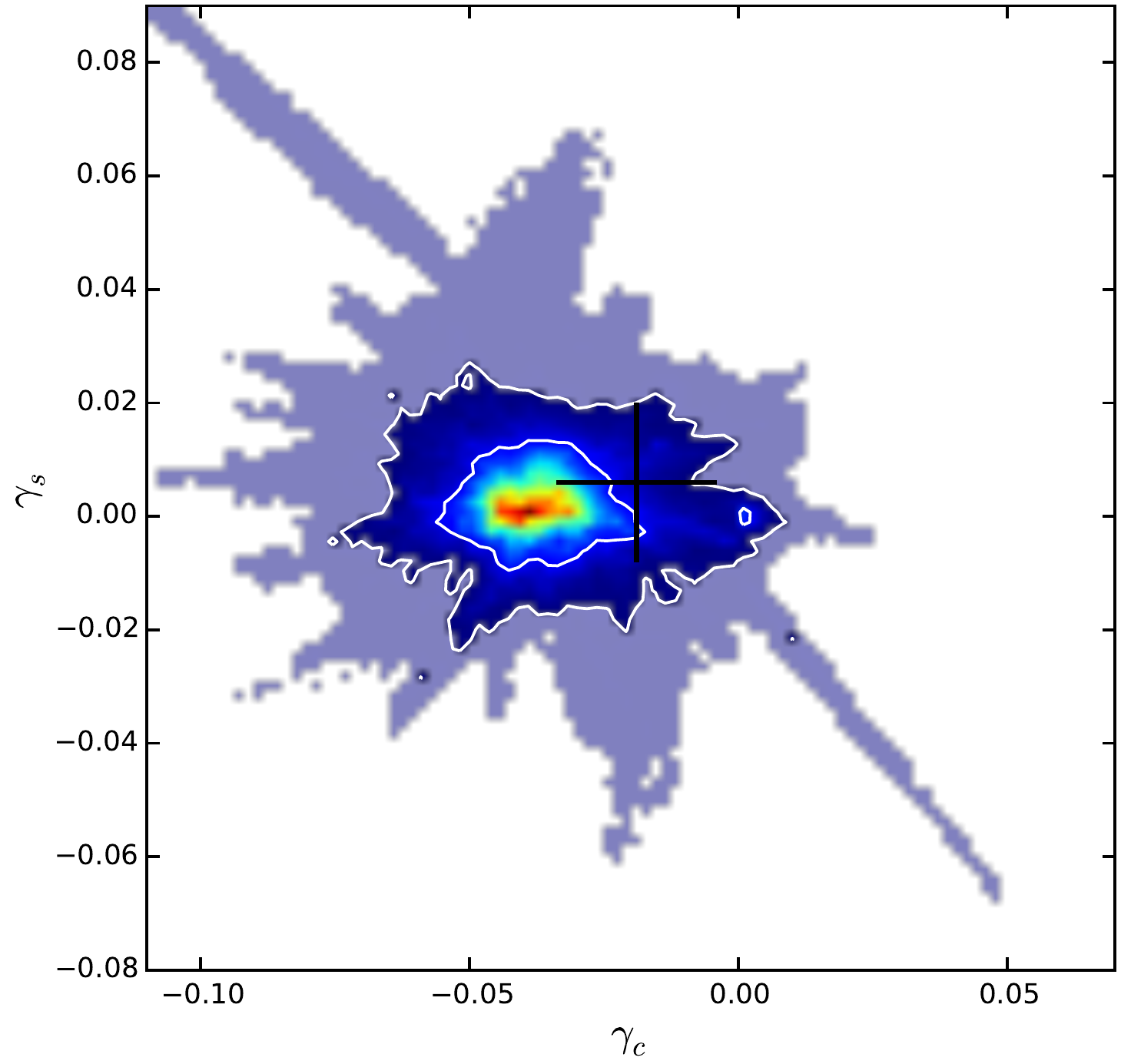}
\caption{Marginalized posterior distribution of fitted external shear values from the Lens+Shear models. $68\%$ and $95\%$ contours are shown in white. The effective shear calculated from the input 3D mass model is shown with a cross. The peak of the distribution from the MCMC models is $\sim 2 \sigma$ away from the truth, similar to what was found by \citet{Wong11}. The offset appears to be inherent to the lens modeling process, possibly due to nonlinear effects from foreground perturbing galaxies.}
\label{fig:shear_compare}
\end{figure}

\citet{Wong11} found that fitted shear parameters from the Lens+Shear models did not always match the shears computed directly from 3D mass models in real lenses, including RXJ1131. We re-examine this result using our controlled simulations. Figure \ref{fig:shear_compare} compares the distributed of fitted shear parameters with the known value of the effective shear for this ENV/LOS. The fitted shear parameters disagree with the true effective shear at the $\sim 2\sigma$ level, broadly consistent with the results found in \citet{Wong11}. While a deeper analysis of the difference is beyond the scope of this work, the results in Figure \ref{fig:shear_compare} suggest that the offsets found by \citet{Wong11} are associated with the lens modeling procedure; they may arise because the fitted shear is attempting to account for nonlinear effects due to foreground perturbing galaxies.

\begin{figure*}[ht]
\begin{center}
\includegraphics[width=1\textwidth]{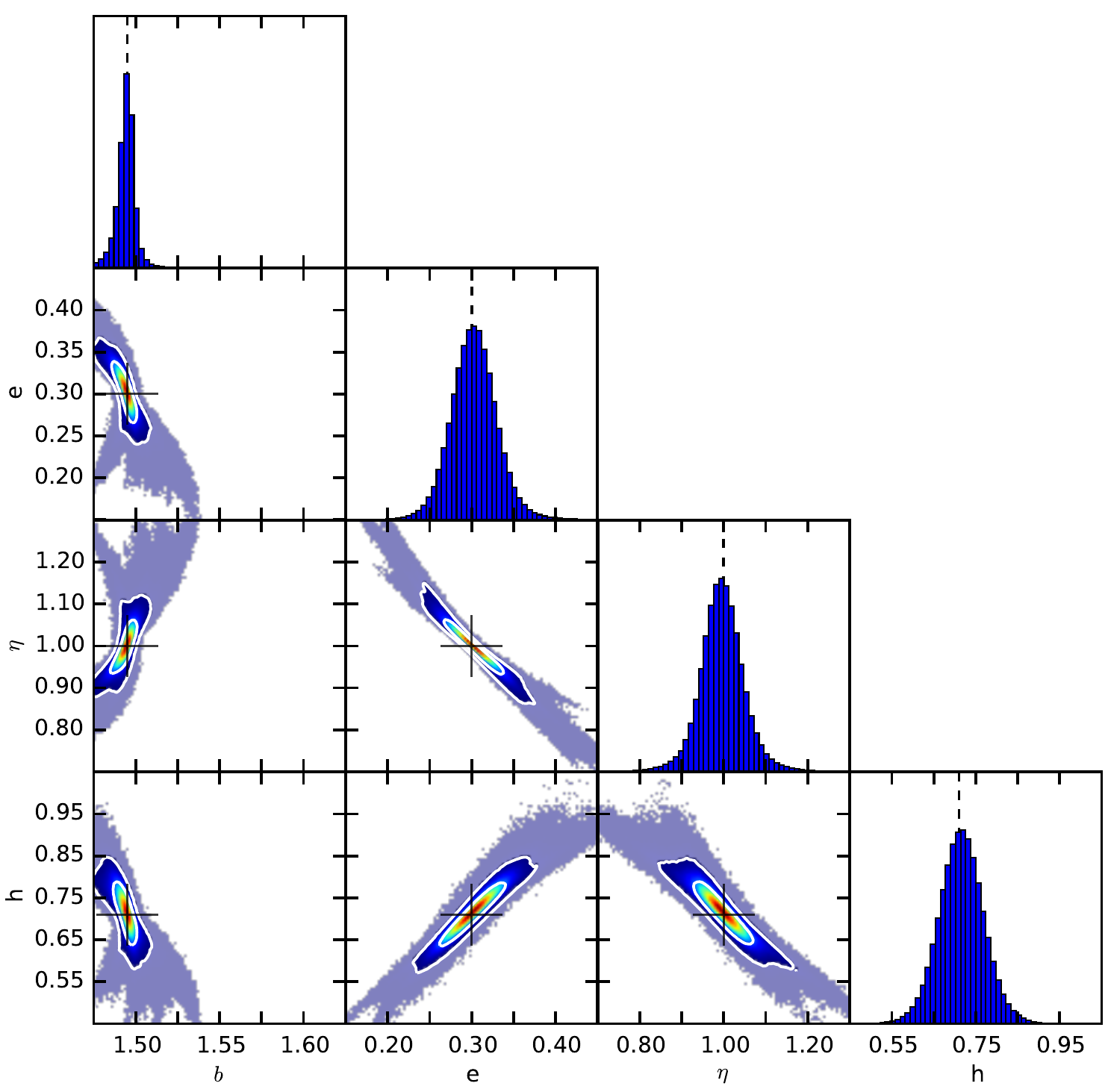}
\caption{\label{fig:los_triangle} Similar to Figure \ref{fig:none_triangle}, but for Lens+3D Tides models. Here all perturbing galaxies are treated with the 3D tidal approximation.  There is no bias in the recovered parameters, implying that omitting higher-order terms will not bias our measurements of the Hubble constant. The scatter in the recovered parameters is much smaller than that of the Lens+Shear and Lens-Only models. However, there are still strong correlations between ellipticity, the power-law index, and the Hubble constant. This is a manifestation of the lens profile degeneracy \citep{Kochanek02}. %
}
\end{center}
\end{figure*}

We now consider our Lens+3D Tides models (Section \ref{sec:3Dtides}). These models include the nonlinear redshift effects, but treat all perturbing galaxies in the tidal approximation. Figure \ref{fig:los_triangle} shows posterior distributions of the recovered lens parameters and Hubble constant. There is no bias in any of the parameters recovered from these models. Our Lens+3D Tides models avoid the biases seen in other models because, rather than treating shear and convergence separately, we build a physical model and then extract both shear and convergence self-consistently. We favor this approach because shear and convergence arise from the same underlying mass distribution and are therefore not independent. 

The main difference between the Lens+Shear and the Lens+3D Tides models is the inclusion of nonlinear effects in the latter. Therefore, the broader posterior distributions that we observe for the Lens+Shear models must be driven by nonlinear effects.

\begin{figure*}[ht]
\begin{center}
\includegraphics[width=1\textwidth]{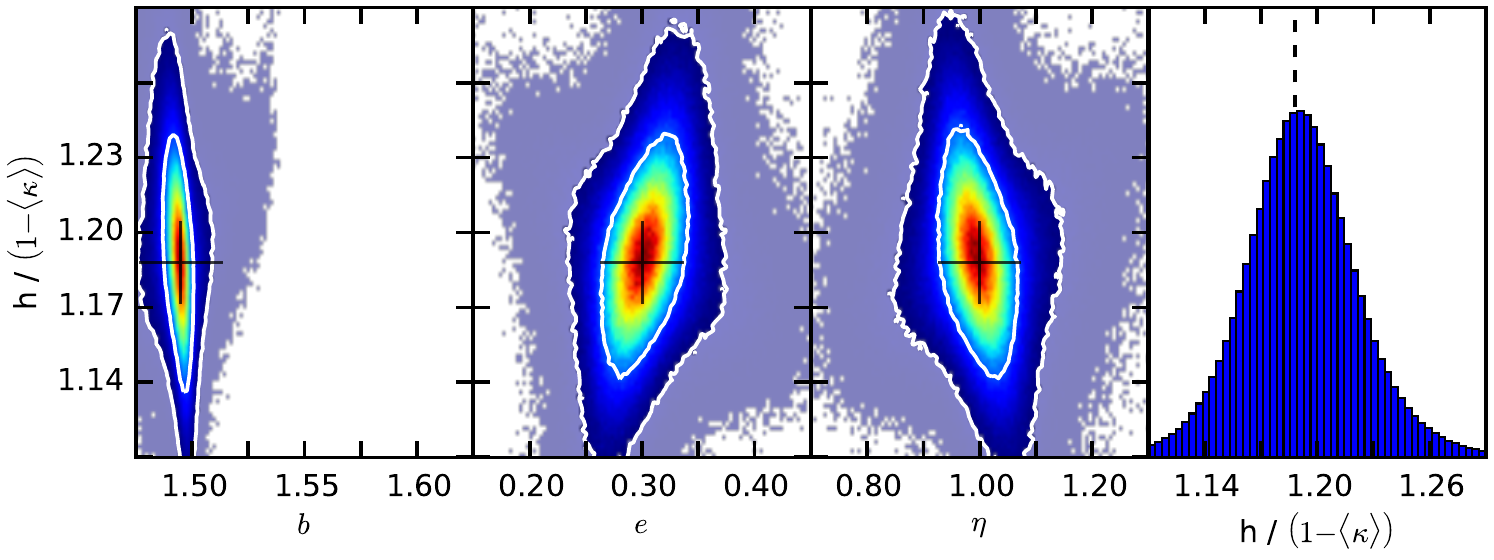}
\caption{\label{fig:scaled_triangle} Recovered parameter distributions for Lens+3D Tides models (see Figure \ref{fig:los_triangle}), but now with the Hubble constant, $h$, scaled to account for the lens profile degeneracy.  We approximate the average convergence at the Einstein radius using equation \ref{eqn:kappa-avg}. The correlations among the Hubble constant, ellipticity, and power-law index are mostly removed. The residual correlations are likely due to higher-order terms in the expression for $h$ \citep{Kochanek02} and the Taylor series expansion for $\langle \kappa \rangle$.%
}
\end{center}
\end{figure*}

However, there are strong correlations among the recovered ellipticity, power-law index, and Hubble constant for our 3D lens models. These correlations involving the Hubble constant are associated with the ``lens profile degeneracy'':  there can be a range of density profiles that all yield a good fit but lead to different values for $h$.  \citet{Kochanek02} shows that to lowest order the degeneracy is characterized by the scaling $h \propto 1 - \langle \kappa \rangle$ where $\langle \kappa \rangle$ is the average convergence from the main lens in the annulus between the image positions.  If we assume that the image annulus is narrow and centered on the Einstein radius, we can evaluate the angle average as
\begin{equation}
\langle \kappa \rangle = \frac{1}{2 \pi} \int_{0}^{2 \pi} \kappa(R_E, \theta) d \theta.
\end{equation}
With our power-law model (eq.\ \ref{eqn:powerlaw}), we can evaluate the integral by making a Taylor series expansion in the ellipticity and using
\begin{equation}\label{eqn:e-series}
\int_0^{2 \pi} \frac{d \theta}{ (1 - 2 e \cos^2(\theta) + e^2\cos^2(\theta))^{1 - \eta / 2}} \approx 1 + e  - \frac{ e \eta}{2}.
\end{equation}
For typical parameter values of interest, we find numerically that this approximation is good to $\sim1\%$.  Then we can approximate the average convergence at the Einstein radius as
\begin{equation}\label{eqn:kappa-avg}
\langle \kappa \rangle \approx \frac{1}{2} \eta  (1 - e)^{2 - \eta} \left(1 +  e - \frac{e \eta }{2}\right).
\end{equation}
We test this analysis by plotting $h/(1-\langle\kappa\rangle)$ in Figure \ref{fig:scaled_triangle}.  Most of the correlation of $h$ with $\eta$ and $e$ has been accounted for.  Remaining correlations are likely due to higher-order terms in the expression for $h$ as a function of profile parameters \citep{Kochanek02} and the Taylor series expansion in equation \ref{eqn:e-series}.

The fact that these correlations are seen in our Lens+3D Tides models suggests that the ENV/LOS does not break the lens profile degeneracy. This is consistent with the idea that ENV/LOS effects are produced by structure on much larger spatial scales than is relevant for the lens profile degeneracy \citep{Schneider13, Xu16}. The lens profile must be broken instead with kinematic information \citep[e.g.,][]{Suyu13} or with extended source modeling \citep{Suyu12}.

\begin{figure}[ht]
\begin{center}
\includegraphics[width=1\columnwidth]{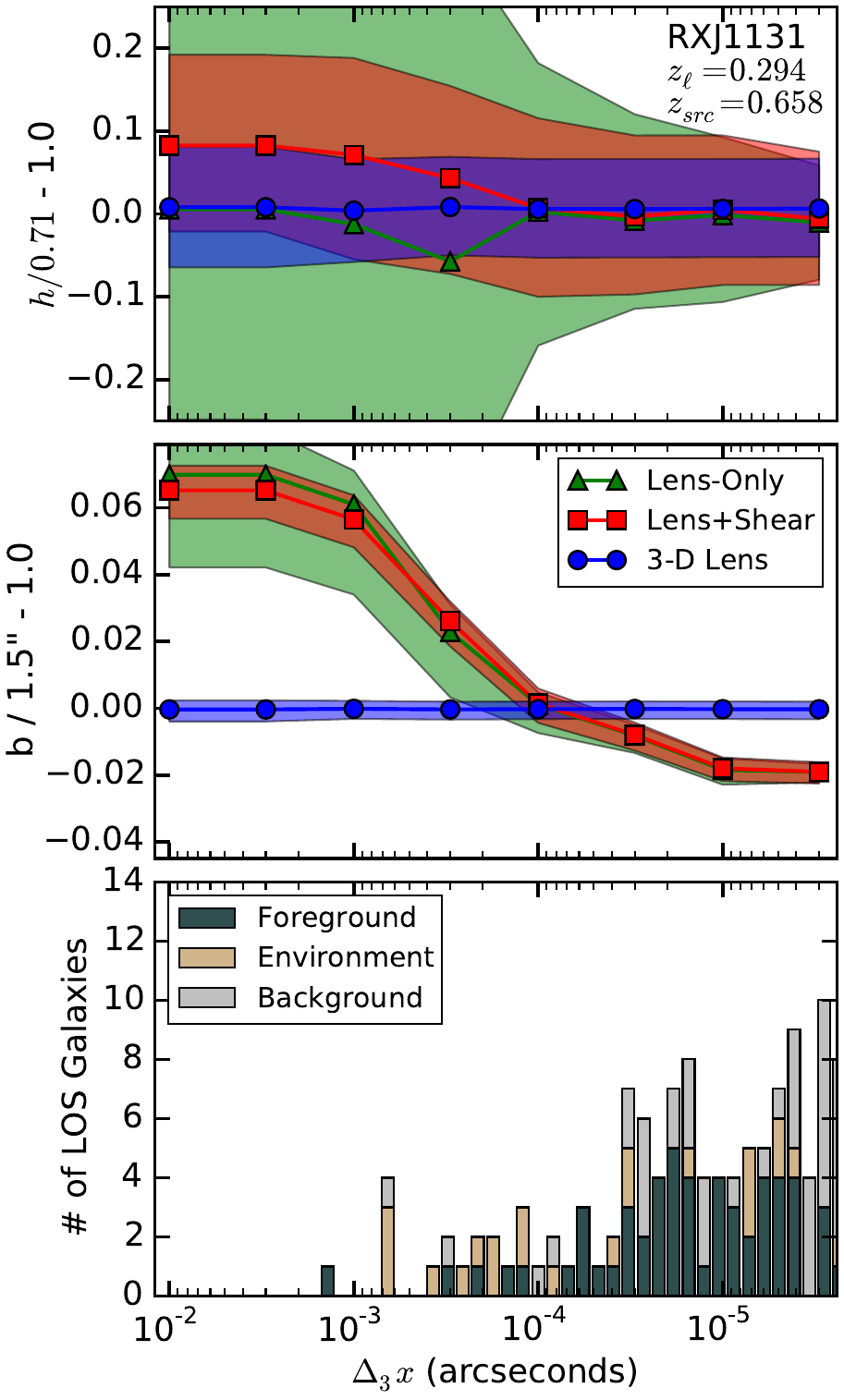}
\caption{\label{fig:RXJ1131} Results for lens models with different thresholds for the tidal approximation, using the RXJ1131 field.  For a given value of $\Delta_3 x$, we run models in which all perturbing galaxies with a larger value of the flexion shift (i.e., those toward the left in the bottom panel) are treated exactly, while perturbing galaxies with a smaller value of the flexion shift (i.e., those toward the right) are treated with the tidal approximation.  Starting at the left, all perturbers are tidal, equivalent to the Lens+3D Tides models (as in Fig. \ref{fig:los_triangle}); moving to the right increases the number of perturbers that are treated explicitly instead.  The top panel shows the median (points) and 68\% confidence (colored band) interval for the Hubble constant recovered from 3D Lens models (blue), Lens+Shear models (red), and Lens-Only models (green). The scatter is driven by the lens profile degeneracy.  The middle panel shows similar results for the Einstein radius parameter. Lens-Only and Lens+Shear models tend to be biased because of external convergence effects which are typically added in post-processing. For RXJ1131, this accounts for a $7\%$ correction, consistent with the peak of the distribution shown in Figure \ref{fig:kappa}. As we include more galaxies exactly, moving to the right of the plot, the bias begins to disappear because we are taking into account more of the convergence explicitly. However, as we take many of the ENV/LOS galaxies into account exactly, the Lens-Only and Lens+Shear models dip below the truth. This is due to the smooth mass correction for voids which is not included in the Lens-Only and Lens+Shear models. Our simulation results suggest that the void correction is $\sim\!2\%$ for RXJ1131.
}
\end{center}
\end{figure}

Finally, we turn to the most general 3D Lens models. Our next goal is to determine when it is valid to use the tidal approximation.  We want to find a sweet spot that speeds up the modeling without introducing systematic uncertainties that are larger than the measurement uncertainties.  Figure \ref{fig:RXJ1131} shows results from 3D Lens model fits with different cuts on the flexion shift, $\Delta_3 x$. Each point on the horizontal axis corresponds to a different \emph{threshold} for $\Delta_3 x$: galaxies with larger flexion shifts are treated exactly, while galaxies with smaller values are treated with the tidal approximation. The leftmost point treats all perturbing galaxies using the tidal approximation and is therefore equivalent to the Lens+3D Tides models. Note that the threshold \emph{decreases} from left to right in the plot; at a given point, galaxies to the left are exact perturbers while galaxies to the right are tidal perturbers.  The error bars mark the $16^{th}$ and $84^{th}$ percentiles of the posterior distribution for the lens model parameters.  Our modeling uses MCMC methods to obtain posterior distributions, so the error bars correspond to marginalized single-parameter posteriors.

The first result from Figure \ref{fig:RXJ1131} is that 3D Lens models do not show any bias in key parameters, even if all of the ENV/LOS galaxies are treated in the tidal approximation, at least for the RXJ1131 field. The scatter in model parameters decreases a little as the $\Delta_3 x$ threshold is reduced and the strongest perturbers are incorporated into the model explicitly. This will be especially important if some extra constraint on the radial profile, e.g., kinematics, is available.  We believe that it is worthwhile to set a conservative threshold of $\Delta_3 x \sim 10^{-4}$ arcsec (i.e., 1.5 dex smaller than our astrometric uncertainties), which amounts to including the strongest 10--15 perturbers explicitly as ``exact'' planes. The computational complexity of the full multi-plane lens equation scales as $\sim N^2$, the reduction from $\sim 300$ galaxies to treating all but 10--15 using the tidal approximation produces an improvement in the performance by a factor of 400--900.

The Lens-Only and Lens+Shear models do show a bias, as mentioned above due to external convergence. Introducing a few of the strongest perturbers (i.e., moving to the right in the figure) means that we begin to explicitly include the most important ENV/LOS galaxies. Thus as we move to the right in the figure, including an ``external'' convergence is no longer necessary because it already included in the models, which in turn reduces the bias in the recovered parameters.  In principle, there is a point at which the sources of convergence are properly accounted for and the bias is removed.  In practice, that point is not easy to determine \emph{a priori}.  Going to very small values of $\Delta_3 x$ (i.e., far to the right in the figure) actually leads to a negative bias.  This is because Lens-Only and Lens+Shear models do not include the void correction discussed in Section \ref{sec:voids}.  This implies that the void correction is  $\sim\!2\%$ in the Einstein radius for the RXJ1131 field.

\begin{figure}[t]
\begin{center}
\includegraphics[width=\columnwidth]{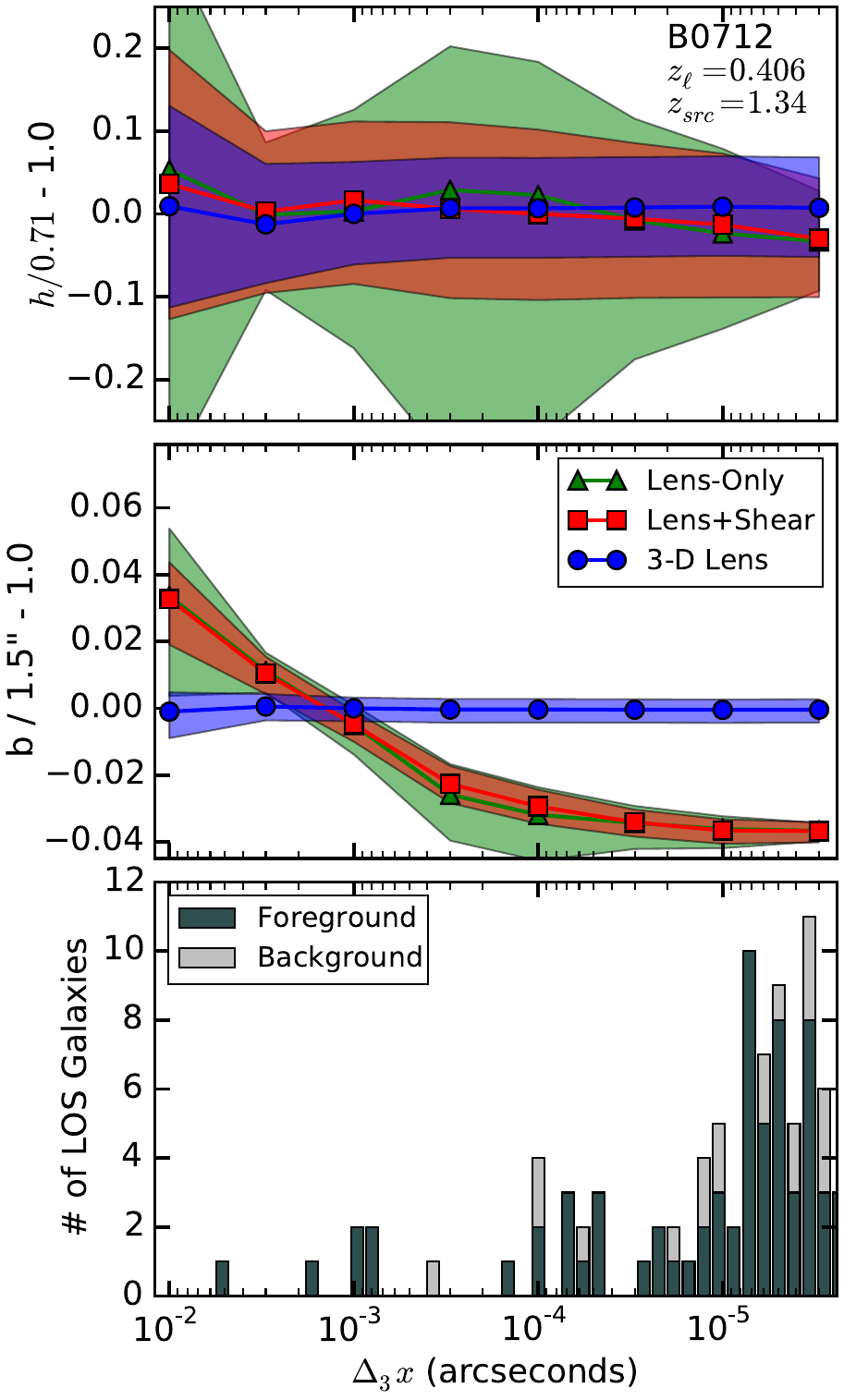}
\caption{\label{fig:B0712} Similar to Figure \ref{fig:RXJ1131}, but for the B0712 field.  This lens is not in a group in our mass models, but it does have a group of galaxies along the LOS.  The qualitative trends from the ENV/LOS effects are similar to what we saw for RXJ1131, but the quantitative details differ. The correction for external convergence (seen at the farthest left point) is smaller for B0712, at $\sim\!4\%$, than what we saw for RXJ1131 at $\sim\!7\%$. This is consistent with the results from Figure \ref{fig:d3xsums} that group lenses typically have stronger ENV/LOS effects than non-group lenses. The correction for the smooth mass density (e.g., voids; inferred from the farthest right point) is larger for B0712 , at $\sim\!4\%$, than for RXJ1131 which only had a correction of 2\%.
}
\end{center}
\end{figure}

So far, we have presented results for a single lens field.  In order to understand whether our conclusions are general, we need to examine how ENV/LOS effects vary from one lensing field to another. One simple test is to follow the full procedure outlined in Section \ref{sec:methods}, using the same fiducial main lens galaxy ($R_E = 1.5\arcsec$ and $e=0.3$) but using a different ENV/LOS.  Figure \ref{fig:B0712} shows results from simulations using the field of B0712+472 (hereafter B0712; \citealt{Jackson98}).  The trends are similar to what we saw for RXJ1131 in Figure \ref{fig:RXJ1131}, but the deviations are somewhat smaller here because the ENV/LOS effects are not as strong in this field (as we saw in Section \ref{sec:Environments}).  The correction due to voids is larger for B0712 at $\sim4\%$. Therefore, the qualitative trends of our results appear to be robust, but the quantitative differences indicate that each lens field needs to be treated on an individual basis.

In this analysis we have placed the same lens galaxy in different mass distributions to see how the ENV/LOS effects vary. The complementary step is to fix the mass in the beam and vary the main lens galaxy to understand how that changes the sensitivity to ENV/LOS effects.

\subsection{What Type of Lens Produces the Strongest Constraint on Cosmology?}
\label{sec:ImageConfigs}

In our final set of simulations (following the procedure in Section \ref{sec:methods}), we choose one ENV/LOS mass distribution (we choose the RXJ1131 field) and test how the constraints on the Hubble constant change when we change the properties of the main lens galaxy and the position of the source behind the lens.

\begin{figure*}
\begin{center}
\includegraphics[width=1\textwidth]{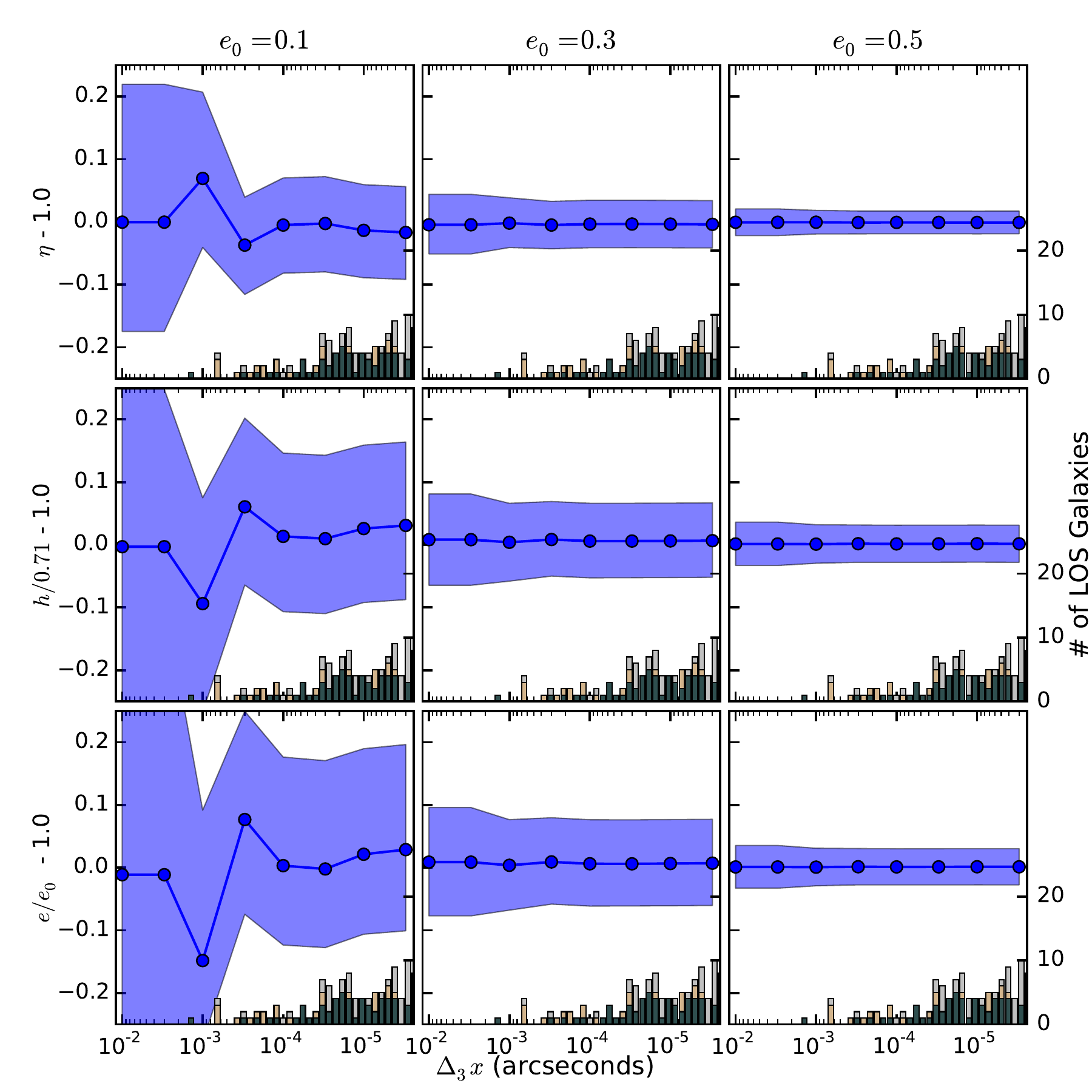}
\caption{\label{fig:ecompare} Recovered model parameters for main lens galaxies with different ellipticities, using the RXJ1131 field. The columns correspond to $e = 0.1,0.3,0.5$ from left to right. Systems with larger $e$ have less scatter in the power-law index, $\eta$, and the Hubble constant, $h$. Elongated lenses tend to produce more asymmetric image configurations. More asymmetric image configurations span a wider range of radii and provide strong constraints on the ellipticity, breaking the lens profile degeneracy and producing tighter constraints on the Hubble constant.%
}
\end{center}
\end{figure*}

One important parameter of the main lens galaxy is the ellipticity. Figure \ref{fig:ecompare} compares results from simulations with different values of $e$.  As the ellipticity increases, the scatter in the recovered Hubble constant decreases.
The image configurations of very asymmetric lens galaxies can only be produced for a smaller range of lens models, leading to stronger constraints on both the ellipticity and the power-law index.  Since those parameters are correlated with the Hubble constant through the lens profile degeneracy, reducing the scatter of fitted $e$ and $\eta$ values leads to a narrower range for $h$ as well.

\begin{figure*}
\begin{center}
\includegraphics[width=1\textwidth]{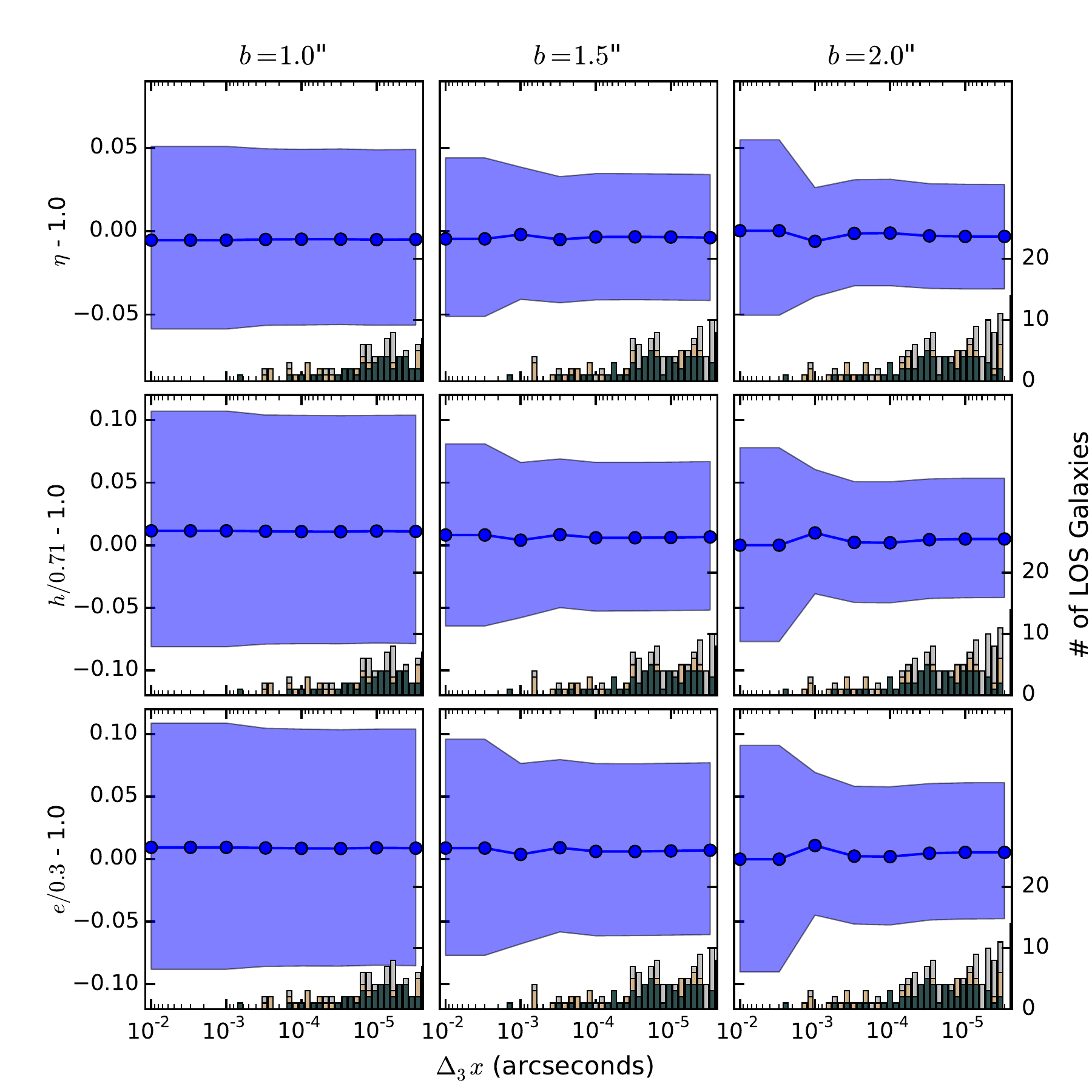}
\caption{\label{fig:recompare} Similar to Figure \ref{fig:ecompare} but for different values of the Einstein radius parameter: $b = 1.0,1.5,2.0$ from left to right. Systems with a smaller Einstein radius have more scatter in the Hubble constant. Constraints on the ellipticity are somewhat stronger for lenses with larger Einstein radii, especially once the few strongest perturbers are included in models. This implies that lenses with a large Einstein radius should produce the strongest constraints on the Hubble constant.%
}
\end{center}
\end{figure*}

A second key parameter of the main lens galaxy is the Einstein radius. Figure \ref{fig:recompare} shows model results for different values of the Einstein radius parameter $b$.  As the Einstein radius increases, the constraints on the Hubble constant get stronger. Lenses with large Einstein radii produce stronger constraints on the ellipticity, limiting the lens profile degeneracy, yielding a tighter distribution on the Hubble constant.

The final lens system parameter we probe is the source position. In previous simulations, we have marginalized over the orientation of the main lens galaxy, which would wash out any trends we would see in the source plane. Instead, we now fix the main lens galaxy to match the real RXJ1131. \citet{Suyu13} find the best-fit parameters of RXJ1131 to be $b = 1.64$\arcsec, $\eta = 1.05$, $e = 0.237$, and $\theta_e = 115.8^{\circ}$. 

\begin{figure}
\begin{center}
\includegraphics[width=1\columnwidth]{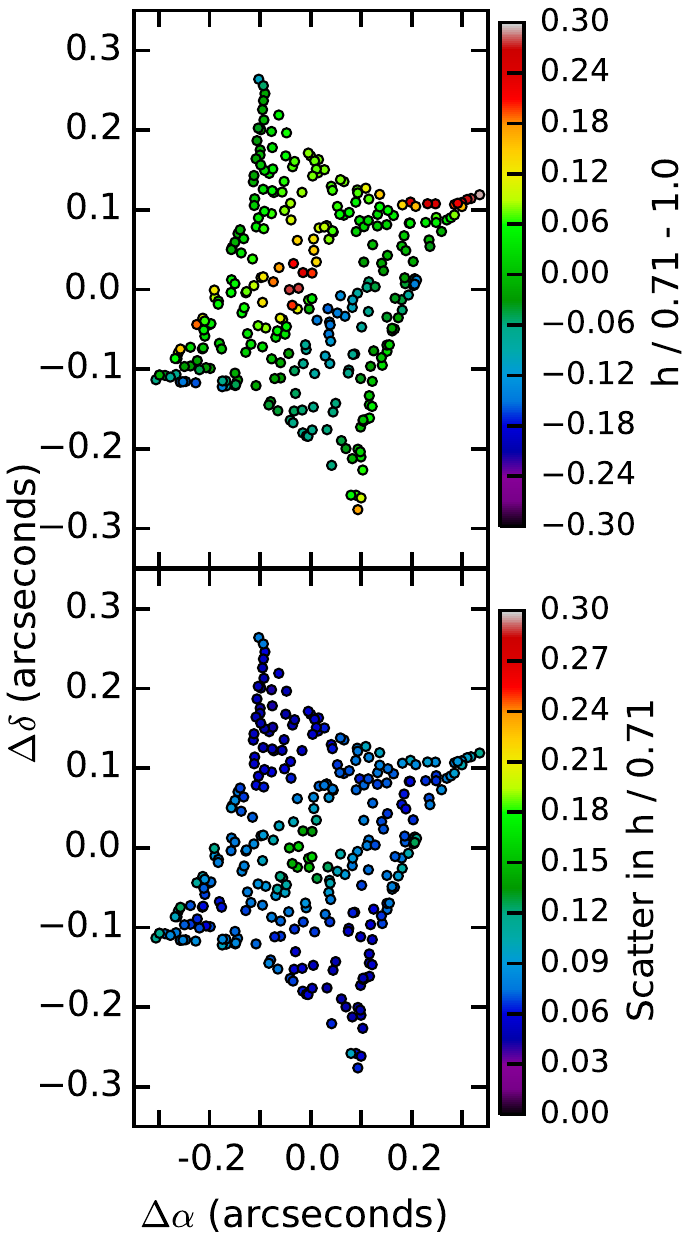}
\caption{\label{fig:srcpos}
Median value (top) and scatter (bottom) for the recovered Hubble constant as a function of source position. The models have little bias and small scatter near caustics, but show an increased scatter/bias near the center. These central source positions correspond to more symmetric image configurations which produce weaker constraints on the ellipticity and are therefore more susceptible to the lens profile degeneracy. More asymmetric image configurations produce stronger constraints on the lens galaxy ellipticity and in turn on the Hubble constant.
}
\end{center}
\end{figure}

Figure \ref{fig:srcpos} shows the median and scatter in the best-fit Hubble constant as a function of source position for the Lens+3D Tides models. The models do well when the source is near a caustic. However, near the center, the models produce larger scatter in the recovered values for the Hubble constant. These source positions correspond to more symmetric image configurations, with the exact center producing an Einstein cross. Symmetric image configurations produce weaker constraints on the ellipticity than asymmetric image configurations and are therefore more susceptible to lens profile degeneracy as discussed above.

One possible approach to using gravitational lensing for precision cosmology is to search for one or a few ``golden lenses'' to use to measure cosmological parameters. One consideration when choosing these lenses needs to be the environment. The results in Figures \ref{fig:ecompare}, \ref{fig:recompare}, and \ref{fig:srcpos} show that asymmetric image configurations, e.g., those produced by a highly elliptical main lens galaxies, are less sensitive to the lens profile degeneracy, leading to improved constraints on the Hubble constant.

\section{Conclusions}

As lensing data improve, it becomes more important to take into account systematic effects like the perturbations due to galaxies in the environment and along the LOS. Our results show that if we want to do ``precision lensing,'' the environment and LOS cannot be ignored.

To understand and quantify how the ENV/LOS affects lensing, we have generated mass distributions based on photometric and spectroscopic observations. We then ray trace through these beams to generate lensing observables. In these calculations, we self-consistently include the contribution from a smooth mass background, for the first time, allowing us to account for mass (in)completeness and voids. We then fit the lensing observables with four types of models: Lens-Only models that ignore the environment, Lens+Shear models that take the common approach of fitting external shear as a free parameter, the Lens+3D Tides models that treat all of the ENV/LOS galaxies using the tidal approximation, but include all nonlinear effects from having mass at different redshifts in the ``3D Tidal Tensors'', and finally 3D Lens models that treat some ENV/LOS galaxies exactly, including higher-order terms.

When considering the effects of individual perturbing galaxies, we find that perturbers in the foreground of the lens affect the lens potential more than those in the background for two different reasons. The first is that background perturbers are downweighted; a background perturber must be closer in projection to have the same effect as a foreground perturber. The second is that while background perturbers can be mimicked by a shear in the lens plane, foreground perturbers create nonlinear effects that cannot be fit with a simple external shear.

Using our results based on individual perturbing galaxies, we define a quantity that we have termed the ``flexion shift,'' $\Delta_3 x$ (eqs.\ \ref{eqn:backgroundd3x} and \ref{eqn:foregroundd3x}), that estimates the deviations in lensed image positions due to third-order (flexion) terms and can be used to rank order the ENV/LOS contributions to the lens potential for an individual perturbing galaxy.

Using this quantity, we find that the importance of environment/LOS effects varies significantly from field to field. Therefore, we argue that each field needs to be modeled individually. Even accounting for the uncertainties in building the 3D mass models, we find that directly calculating the external convergence from the 3D mass models produces a narrower distribution than that from ray tracing through $N$-body simulations. This translates to a stronger prior on measuring the Hubble constant. We find that lens galaxies that are in groups tend to have a stronger contribution from the ENV/LOS than those that are not in groups. 

We show that fitting lens models that ignore the ENV/LOS (Lens-Only) does not reproduce the input lens system parameters or the Hubble constant. Models that fit an external shear, the Lens+Shear models, overpredict the Hubble constant and the Einstein radius of the main lens galaxy; these quantities need to be corrected using some other constraint on the external convergence, which is typically done in post-processing. The Lens+3D Tides models do not produce this bias (or biases in any other model parameters) because we explicitly include the convergence in the lens models, implying that higher-order terms do not lead to biases in the derived cosmological parameters. It is still possible that there may be some unexpected bias from how we construct our mass distributions, but that is beyond of the scope of this paper and will be considered in more detail in future works. 

We show that all models, even when we include all galaxies exactly, are fundementally limited by the lens profile degeneracy in agreement with \citet{Xu16}, \citet{Schneider13}, and  \citet{Sluse12}. Extra information like kinematic measurements or extended source reconstruction techniques are necessary.

When using our 3D Lens model, we still must choose which galaxies are treated exactly and which are treated in the tidal approximation. Our tests suggest that a conservative cutoff in $\Delta_3 x$ is $10^{-4}$ arcsec, a factor of 30 smaller than the assumed uncertainties on the observed positions. This cut typically requires 10-15 galaxies of $\sim 300$ galaxies within 5 arcmin to be treated exactly, making the calculation $\sim 400-900$ times more efficient than the full multi-plane lens equation.

LSST and Euclid will find an immense number of new strong lens systems. There are various strategies about how to use this upcoming data set for cosmology. One possibility is to use all of the lenses to beat down uncertainties using statistics. However, if we are entering the systematics-limited regime, this approach will succeed only if one can account for systematic uncertainties (e.g., with our 3D Lens models). An alternative strategy is to use the large number of lenses discovered by LSST and Euclid to search for a few rare, ``golden'' lenses that have small systematic uncertainties. One possible criterion for a ``golden'' lens could be to have a weaker contribution from the ENV/LOS. Based on our analysis, to minimize ENV/LOS effects, we should search for lenses with large Einstein radii and asymmetric image configurations as these will be less sensitive to the lens profile degeneracy. While suggestive, these results merit further investigation to get the most out of future surveys like LSST and Euclid.

Throughout this work we have assumed that our lensing beams perfectly describe the ENV/LOS, but there is also uncertainty in generating the mass distributions that is only briefly addressed here. This source of uncertainty will be explored in a forthcoming paper (K.\ C.\ Wong et al. 2017, in preparation).

\acknowledgements
We thank the anonymous referee for his/her comments and careful consideration. We thank Edi Rusu, Phil Marshall, Sherry Suyu, Chris Fassnacht, and Stefan Hilbert for enlightening discussions about 3D lensing. We thank Iair Arcavi and Griffin Hosseinzadeh for discussions about presentation of these results. C.V.M acknowledges support from NSF grant AST-1313484. C.R.K acknowledges funding from NSF grant AST-1211385. A.I.Z acknowledges funding from NSF grant AST-1211874 and NASA grant ADP-10AE88G. She also thanks the John Simon Guggenheim Foundation and the Center for Cosmology and Particle Physics at NYU for their support. K.C.W is supported by an EACOA Fellowship awarded by the East Asia Core Observatories Association, which consists of the Academia Sinica Institute of Astronomy and Astrophysics, the National Astronomical Observatory of Japan, the National Astronomical Observatories of the Chinese Academy of Sciences, and the Korea Astronomy and Space Science Institute.

\bibliographystyle{aasjournal}
\bibliography{sims}

\end{document}